\RequirePackage[2020-02-02]{latexrelease}
\documentclass[twocolumn, superscriptaddress,aps, amsmath, amssymb, prl,letter]{revtex4-2}
\usepackage{color}
\usepackage{dcolumn}
\usepackage{graphicx}
\usepackage{subfigure}
\usepackage{amsmath}
\usepackage{comment}
\usepackage{todonotes}

\begin{document}
	
\title{The excitatory-inhibitory branching process: a parsimonious
 view of cortical asynchronous states, excitability, and criticality.}

 \author{Roberto Corral L\'opez}
\affiliation{Departamento de Electromagnetismo y F{\'\i}sica de la Materia and Instituto Carlos I
de F{\'\i}sica Te{\'o}rica y Computacional. Universidad de Granada.E-18071, Granada, Spain}
\author{V{\'\i}ctor Buend{\'\i}a} 
\affiliation{Department of Computer
  Science, University of T{\"u}bingen, and Max Planck Institute for
  Biological Cybernetics, T{\"u}bingen, 72076, Germany}
\author{Miguel A. Mu\~noz} 
\affiliation{Departamento de Electromagnetismo y F{\'\i}sica de la Materia and Instituto Carlos I
de F{\'\i}sica Te{\'o}rica y Computacional. Universidad de Granada.E-18071, Granada, Spain}

	\begin{abstract}
          The branching process is the minimal model for propagation
       dynamics, avalanches and criticality, broadly used in
          neuroscience.  A simple extension of it, adding inhibitory
          nodes, induces a much-richer phenomenology, including, an
          intermediate phase, between quiescence and saturation, that
          exhibits the key features of ``asynchronous states'' in
          cortical networks.  Remarkably, in the inhibition-dominated
          case, it exhibits an extremely-rich phase diagram, that
          captures a wealth of non-trivial features of spontaneous
          brain activity, such as collective excitability, hysteresis,
          tilted avalanche shapes, and partial synchronization,
          allowing us to rationalize striking empirical findings
          within a common and parsimonious framework.
	\end{abstract}

	\maketitle
        The idea that information-processing systems, both 
          biological and artificial, can extract important functional
        advantages from operating near the edge of a phase transition
        was already suggested by A. Turing in 1950, inspiring since
        then theory and experiment \cite{Mora,RMP}. Beggs and Plenz,
        pioneering the experimental search for signatures of
        criticality in neural systems, found scale-free outbursts of
        neuronal activity occurring in between consecutive periods of
        quiescence, i.e., \emph{neuronal avalanches} \cite{BP2003}, as
        consistently reported across brain regions, species, and
        observational scales \cite{BP2003,Petermann09, Haimovici,
          Taglia,Plenz-Shriki2013,2-photon,Plenz-task}.  These
          avalanches have sizes and durations distributed as power
        laws with exponents consistent with those of a critical
        branching process (BP) \cite{Liggett,Harris} and often
          exhibit a parabolic shape on average (another trademark of
          critical BPs) \cite{Sethna,shape,Plenz2021,Serena-BP}.  In
        spite of some methodological caveats
        \cite{Viola1,neutral,Viola2}, experimental discrepancies
        \cite{Fontenele}, and the existence of alternative
        interpretations \cite{Crackling,LG,hybrid,Suweis}, the
        empirical observation of scale-free neuronal avalanches
        triggered renewed interest in the idea of criticality in brain
        networks \cite{Schuster,Chialvo2010,Levina,Beggs2021} and its
        potential relevance for computation and information processing
        \cite{Kinouchi,shew_optimal, Plenz-functional,Turrigiano} (see
        also
        \cite{Plenz-review,Mora,RMP,Breakspear-review,Viola-review}).
      
  \begin{figure}
	\includegraphics[width=0.45\textwidth]{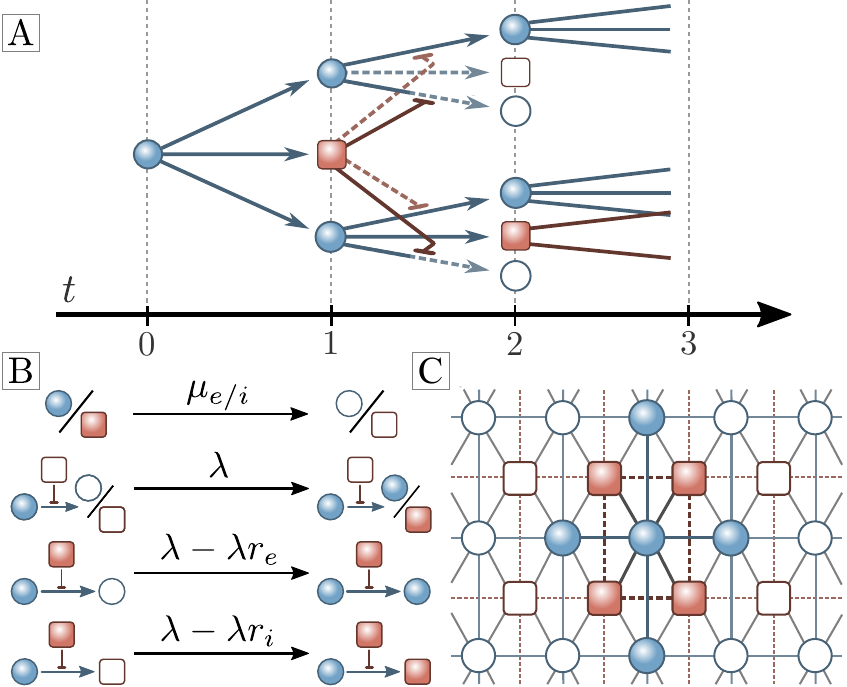}
        \caption{(A) Sketch of an excitatory-inhibitory branching
          process on a tree; active inhibitory units (red squares)
          reduce the probability of propagation from active excitatory
          units (blue circles).  Empty symbols stand for inactive
          units and full/dashed lines for fulfilled/unfulfilled
            processes.  (B) Transition rates for the
          ``excitatory-inhibitory contact process'' (EI-CP). (C)
          Illustration of a two-dimensional (2D) lattice with a
          central cluster of active nodes. }
\label{fig:Sketch}
\end{figure}
Nevertheless, the stylized picture of neuronal activity as a BP seems
exceedingly na{\"i}ve, as it overlooks the fact that about $20 \%$ of
the neurons in the cortex are inhibitory ones \cite{Dayan} and that these
play a crucial role in shaping cortical activity
\cite{Scanziani,Doiron,Deneve}. Actually, the ``standard model'' of
spontaneous brain activity is that of a ``\emph{balanced state}'' in
which excitatory and inhibitory inputs to any given neuron nearly
cancel each other on average, giving rise to a fluctuation-dominated
``\emph{asynchronous state}'' \cite{vVS,Brunel,Brunel2,Sompo1}. This is
characterized
        by rather irregular (Poisson-like) single-neuron activations,
        delayed correlations between excitation and inhibition, and
        small averaged pairwise correlations, etc. 
        \cite{Renart,Yuste,Scanziani,Helias}.
        These properties, important for efficient encoding of
        information \cite{Renart,Scanziani,Deneve,Doiron}, are
        markedly different from those of usual critical states but are
        also crucial for information processing, suggesting that
        critical and asynchronous states could act complementarily to
        tackle diverse functional tasks (e.g., requiring either strong
        correlation for collective response or decorrelation to limit
        redundancy). Hence, describing these alternative states under
        a common overarching framework is a timely and challenging
        goal \cite{Benayoun,Helias,Zierenberg18,Zhou-2020,Buendia-Jensen,
          Girardi, Li-Shew}.

        Here, we analyze what happens in archetypical models of
        activity propagation ---such as the BP or, more specifically,
        its continuous-time counterpart: the contact process
        \cite{Liggett,Harris}--- if, as sketched in
        Fig.~\ref{fig:Sketch}, inhibitory units are considered in
        addition to the usual excitatory ones? Do additional phases beside
        the standard ``quiescent'' and ``active'' ones emerge?
        \cite{Marro,Henkel,Odor}.  What are their key
        features and phase transitions?  In what follows we answer
        these questions, elucidating an extremely rich phenomenology
        that reproduces the key features of ``asynchronous states'',
        but also collective excitability, bistability, non-parabolic
        avalanches, quasi-oscillations, criticality, etc., allowing us
        to rationalize a wealth of striking empirical observations in
        a parsimonious way.

\begin{figure*}
  \includegraphics[width=0.9\textwidth]{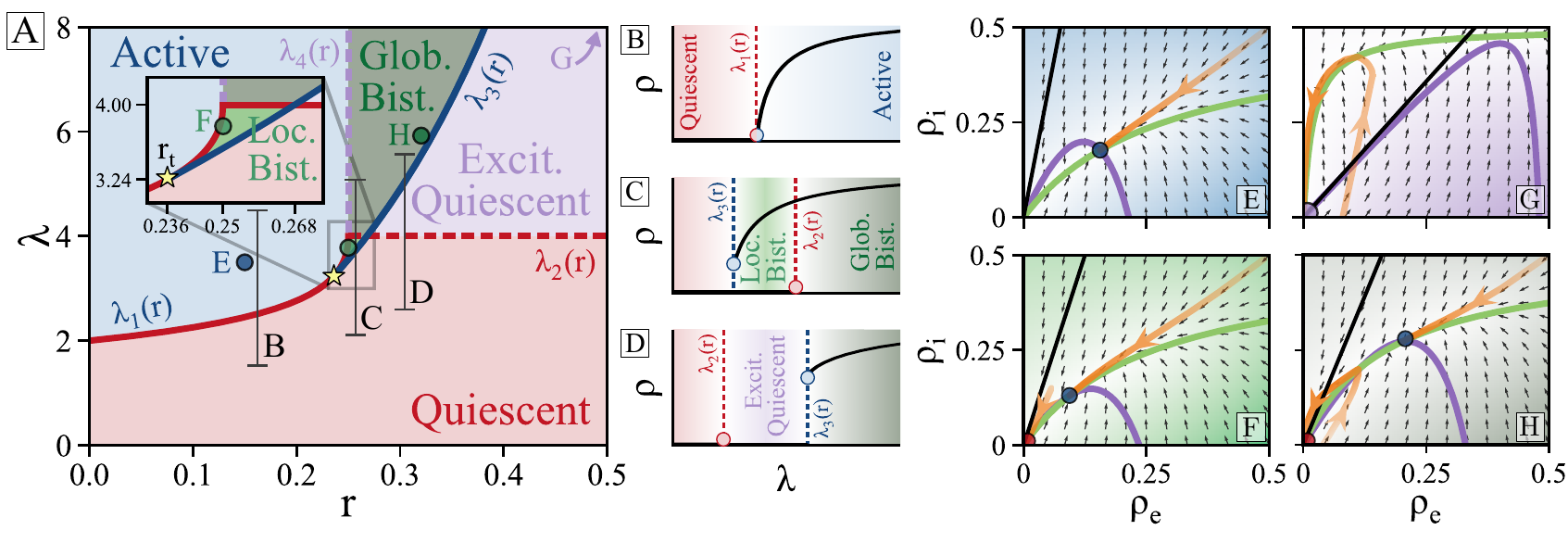}
  \caption{Results for the excitatory-inhibitory contact process
    (EI-CP) on fully-connected networks as analytically obtained from
    Eq.(\ref{Eqn:MF}) for $\alpha=1/2$, $r_i=0$, and
    excitation-dominated initial conditions (note that
    inhibition-dominated conditions always lead to the quiescent
    state).  (A): Phase portrait in the $r$-$\lambda$ plane: active
    phase (blue), quiescent phase (red), \emph{excitable quiescent}
    phase (purple) and bistable regimes (green).  The full red line
    $\lambda_1(r)$ (resp. $\lambda_3(r)$ in blue) marks continuous
    (resp. discontinuous) transitions between quiescent an active
    states. These two lines come together at a tricritical point
    (yellow star).  The line $\lambda_2(r)$ marks a Hopf bifurcation,
    separating the standard quiescent phase from a excitable quiescent
    one, where the quiescent state is locally unstable, but globally
    stable.  (B-D) Overall stationary activity $\rho=\rho_e+\rho_i$ as
    a function of $\lambda$ for three different values of $r$ as
    marked and color-coded in (A): continuous transition (B),
    discontinuous transition with a regime of bistability between an
    active state and a quiescent state (C) or between an active
      and an excitable quiescent state (D).  (E-G): Flow diagrams in
    the $\rho_e, \rho_i$ plane for the three points marked in panel
    (A); the background color stands for the phase and its color
    intensity is proportional to the vector-field module, the colored
    lines are the nullclines: $\dot{\rho}_e=0$ (green) and
    $\dot{\rho}_i=0$ (purple) respectively, and the black line
    ($\rho_i=\rho_e/r$) separates zone 1 (inhibition dominated) from
    zone 2 (excitation dominated).  Characteristic trajectories are
    depicted as arrowed orange lines, while colored points stand for
    stable steady states.  }
  \label{fig:MF}
\end{figure*}

The excitation-inhibition contact process (EI-CP) is a generalization
of the ordinary contact process (CP)
\cite{Harris,Liggett,Marro,Henkel,Odor}, operating on
top of an arbitrary directed network in which active
excitatory neurons attempt to propagate activity to their neighbors,
while inhibitory ones hinder such a propagation \cite{Dayan}.  We
consider diverse types of network architectures, such as fully-connected graphs,
sparse random networks, and two-dimensional (2D) lattices (see
Fig.\ref{fig:Sketch}C).

The networks consist of $N$ nodes, of which a fraction $\alpha$
  are excitatory (E) and the remaining $(1-\alpha)N$, are inhibitory
  (I), a proportion that is preserved for the inward
  connectivity of every single node.  The state of each node $j$ at
time $t$ is defined by a binary variable ($s_j(t)=1$ for active nodes
and $s_j(t)=0$ for inactive or ``silent'' ones), and $\rho_{e}(t)$
(resp. $\rho_{i}(t)$) is the fraction of active excitatory
(resp. inhibitory) nodes. The dynamics is akin to the ordinary CP:
active nodes become silent at a fixed rate $\mu_{e}=\mu_i=1$, but only
active \emph{excitatory} nodes can propagate activity to each of their
silent nearest neighbors at a rate $\lambda/K$. On the other hand,
each active inhibitory node reduces the rate at which each neighbor is
activated by $r_{e/i} \lambda/K$ (for E/I units, respectively), with
$0 \leq r_{e/i} \leq 1$ (Fig. \ref{fig:Sketch}A). Thus, the activation
rate of a silent node $j$ is
$ f\left(\frac{\lambda}{K}\sum_{k\in\Omega_j^e}s_k -\frac{\lambda
    r_{(e/i)}}{K}\sum_{k\in\Omega_j^i}s_k\right)$, where
$\Omega_j^{e/i}$ is the set of E/I neighbors of node $j$ in the
considered network, and the gain function,
$f(\Lambda) = \max(0, \Lambda)$, enforces the non-negativity of the
transition rates. We focus on the \emph{asymmetric} variant of the
model, in which inhibition acts more strongly on excitatory than on
inhibitory nodes, i.e.  $r\equiv r_e>r_i$.  leading to
\emph{inhibition-dominated} networks.  For simplicity, here we fix
$r_i=0$ ---i.e. no inhibition to inhibitory nodes--- and $\alpha=1/2$
(see SI for generalizations). The Master equation defined by the above
rates can be integrated in an exact way with Gillespie's algorithm
\cite{Gillespie} and also studied analytically (see SI).

Let us first discuss the case of fully-connected networks, for which
mean-field equations (exact in the infinite-$N$ limit) can be derived
from a standard size expansion \cite{Marro,Henkel,Romu},
\begin{eqnarray}\label{Eqn:MF}
  \dot \rho_e (t)&=& -\rho_e +~~ (\alpha-\rho_e) ~ f(\lambda(\rho_e -
r\rho_i)), \nonumber \\ \dot \rho_i (t) &=& -\rho_i +
   (1-\alpha-\rho_i) ~ f(\lambda  \rho_e),
\end{eqnarray}
while for finite $N$, additional (demographic) noise terms need to be
added to Eqs.(\ref{Eqn:MF}) (see SI).  Notice that Eq.(\ref{Eqn:MF})
is a version of the celebrated Wilson-Cowan model for neural
dynamics \cite{WC-review,Negah}, and that, actually, our full
  model is also a variant of the ``stochastic Wilson-Cowan model'',
  for which many illuminating results have been obtained in the
  \emph{symmetric case},
  \cite{Benayoun,Serena-NN,Lucilla,Lucilla2}. However, here we focus
  on the inhibition-dominated \emph{asymmetric case}, which exhibits a
  much richer phenomenology (see below).

\begin{figure}
\includegraphics[width=0.45\textwidth]{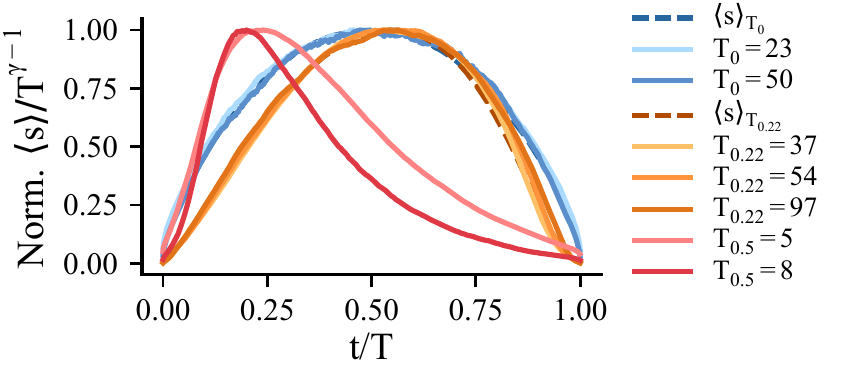}
\caption{Avalanche shapes rescaled with their duration $T_r$ at
    different points of the mean-field phase diagram (characterized by
    $r$). Simulations are performed for the noisy version of
  Eq.(\ref{Eqn:MF}) with small excitatory initial conditions (see
  SI-IV.B).  (i) At the inhibition-free critical point (blue curves;
  $r=0$, $\lambda_c=2$) avalanches for different durations, $T_{r=0}$,
  are scale-free as their rescaled curves collapse onto a universal
  inverted-parabola shape using the BP exponent $\gamma=2$
  \cite{shape,Serena-BP}. (ii) In the presence of inhibition, the
  curves at the critical point (orange curves; $r=0.22$ and
  $\lambda_c=3$) are scale-invariant with BP exponents and they
  collapse onto a slightly ``tilted'' non-parabolic curve (see also
  \cite{Lucilla}). (iii) Within the excitable phase (red curves;
  $r=0.5$, $\lambda=10$), i.e.  away from bifurcations, one
  observes duration-dependent (non-scale-invariant) skewed non-parabolic
  shapes.}\label{Fig:avalanches}
 \end{figure}

Observe that, owing to the piecewise
    definition of $f$, Eq.(\ref{Eqn:MF}) is a non-smooth dynamical
    system \cite{Kunze} and the space of states ($\rho_e,\rho_i$) is
    divided in: (i) a \emph{zone 1}, with $\rho_e - r\rho_i <0$, for
    which the gain function in the equation for $ \dot \rho_e (t)$
    vanishes so that the quiescent state $\rho_e=\rho_i=0$ is always
    reached, and (ii) a \emph{zone 2}, for $\rho_e - r\rho_i >0$ which
    ---as shown in Fig.\ref{fig:MF}--- entails a rich phase diagram
    including a quiescent phase, an active one, and a regime of
    bistability (the corresponding nullclines, fixed points and,
    characteristic trajectories are shown in Fig.\ref{fig:MF}E-H; see
    also \cite{WC-review,Izhi,Negah}). Observe that the transition from
    active to quiescent can be either (i) continuous, as in the
    standard CP (line of transcritical bifurcations at
    $\lambda_{1}(r) = \frac{4}{1+\sqrt{1-4r}}$ for $r\leq 1/4$; red
    line in Fig.\ref{fig:MF}A), (ii) discontinuous with bistability,
    (saddle-node bifurcations at $\lambda_{3}(r)=\frac{8r}{(r-1)^2} $;
    blue line in Fig.\ref{fig:MF}A), or (iii) \emph{tricritical} at
    their merging point ($r_t=\sqrt{5}-2$; yellow star). Note
    also the presence of a line of Hopf bifurcations
    ($\lambda_{2}(r) \equiv 4$ for $r \geq 1/4$; red horizontal dashed
    line) where the quiescent state loses its local stability,
    suggesting the emergence of oscillations above it. However, the
    non-smoothness of the dynamical system leads to \emph{frustrated
      oscillations}, i.e. excitatory perturbations (in zone 2)
      give raise to curved trajectories that cross to states in zone 1
      and then decay back to quiescence (see Fig.\ref{fig:MF}G).
    This generates an ``\emph{excitable phase}'' above the Hopf line
    where the quiescent state is \emph{locally unstable} to excitatory
    perturbations, so that these can be hugely amplified before
    relaxing back to quiescence, making it \emph{globally
      stable}. This creates a mechanism for bursting/avalanching
      behavior, related but different from the one studied in
      \cite{Benayoun,Serena-NN}.  This type of
    transient-amplification effect is well-known to stem from the
    non-normal (non-Hermitian) form of the Jacobian matrix
and its concomitant non-orthogonal eigenvectors
and its implications have been long studied in neuroscience
  \cite{MurphyMiller,Gertsner,WC-review,Benayoun}.
A particularly interesting case of non-normality occurs where the
transcritical and Hopf lines meet, i.e.  at the codimension-2
Bogdanov-Takens (BT) bifurcation \cite{Izhi},
characteristic of, so-called, \emph{non-reciprocal phase
  transitions}, a currently hot research topic \cite{Non-reciprocal}.
The non-normal nature of the dynamics entails a number of non-trivial
features such as tilted avalanches ---characterized by a highly non-parabolic
averaged shape as shown in Fig.\ref{Fig:avalanches}--- which appear
all across the excitable phase when excitatory inputs perturb the
quiescent state. Note that they are not scale invariant, i.e., they have
diverse, duration-dependent, shapes (see also \cite{Lucilla}).  It is
only at the line of continuous transitions that avalanches are both
tilted and scale-free, resembling the non-parabolic scale-free
avalanches reported in, e.g., zebra-fish experiments
\cite{zebra}. Avalanches become parabolic only when inhibition is
switched off ($r=0$) and their scaling differs from the standard BP
only at the exceptional BT point (its ``exotic'' critical features
will be scrutinized elsewhere \cite{Helena}).  Thus, in summary, the
asymmetric (inhibition-dominated) EI-CP model exhibits a much-richer
phenomenology than its standard CP counterpart (and that the symmetric
version of the model, see SI) already at a mean-field level.

\begin{figure*}
  \includegraphics[width=0.95\textwidth]{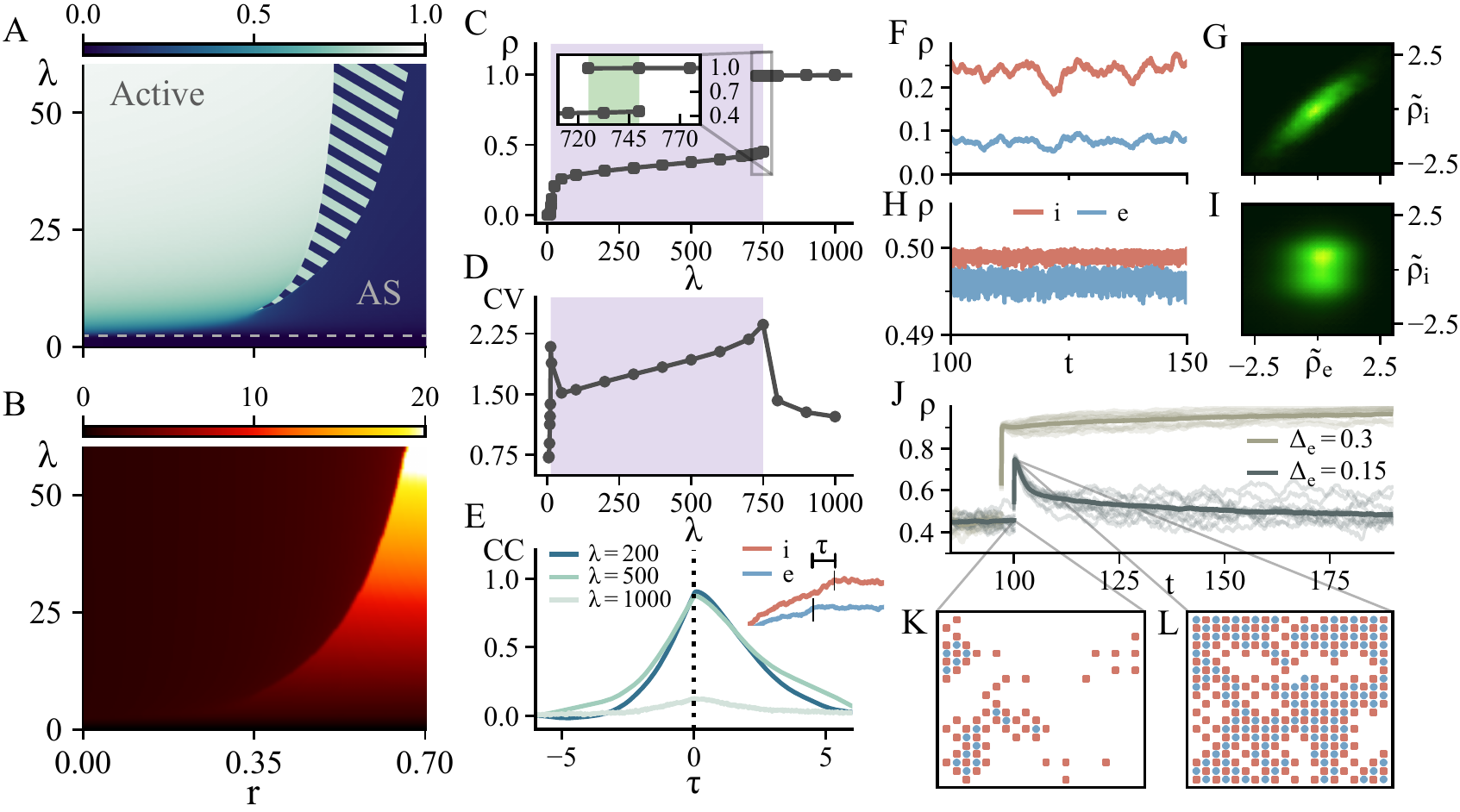}
  \caption{Features of the asynchronous phase. (A-B) Analytical
    results for sparse (annealed) random networks: (A) Stationary
    activity where the dashed line indicates the end of the quiescent
    phase and stripes signal the bistability region between the
    asynchronous and standard-active phase. (B) Henrici index gauging
    the level of non-normality in the $r,\lambda$ diagram (see
    SI-IV.C). (C-L) Results for a 2D lattice (which helps
    visualization) with $N=10^4$. (C) Section of the phase diagram
    ($r=0.7$), illustrating the discontinuous transition with
    bistability and (D) coefficient of variation ($CV$) for different
    values of $\lambda$. (E) Lagged cross-correlations ($CC$) between
    excitatory and inhibitory time series; inhibition follows closely
    excitation with a delay $\tau$. Total excitatory and inhibitory
    activity as a function of time for $r=0.7$ is plotted in (F) for
    the AS phase ($\lambda=200$) and in (H) for the standard active
    phase ($\lambda=1000$); (G-I) same as in F and H, respectively,
    plotted in the standarized $\tilde\rho_e, \tilde\rho_i$ plane (see
    SI-IV.C), as an illustration of the diverse nature of
    cross-correlations in both phases.  (J) Stimulation experiment
    where a fraction $\Delta_e$ of excitatory nodes (on a 2D
      lattice) is transiently activated; in the bistability region
    ($r=0.7$, $\lambda=750$), this can potentially drive the system
    from the AS phase to the standard active phase, much as in
    experimental setups \cite{Yang-Dan}. (K/L) Snapshots of the system
    before (K) and after (L) the perturbation ($N=20^2$). See SI for
    further details.}\label{fig:AS}
	\end{figure*}
        To go beyond mean-field, we now study sparse networks
        ($K \ll N$), and scrutinize the effects of their inherent
        stochasticity.  In particular, we start by considering
        analytically-tractable \emph{annealed random networks} ---in
        which the $\alpha K$ excitatory and $(1-\alpha) K $ inhibitory
        neighbors of each single node are randomly selected at each
        time step.  In this way, the input to each neuron is a random
        variable, whose probability distribution can be
        straightforwardly seen to be the product of two binomials (see
        SI).  From this probability distribution, one can then compute
        the mean activation rate for each node,
        $\langle f (\rho_e,\rho_i) \rangle_K$, which ---as a
        consequence of Jensen's inequality \cite{Buendia-Jensen}---
        turns out to be larger than its mean-field counterpart
        $f (\langle \rho_e \rangle_K, \langle \rho_i \rangle_K)$, in
        Eq.\ref{Eqn:MF}.  The resulting exact equation can be solved
        using series expansions or numerically (see SI).  The most
        salient feature of its associated phase diagram (Fig.4A)
        is the emergence of an intermediate phase between the
        standard quiescent and active phases. It is separated from the
        former by a line of continuous transitions
        ($ \lambda_c(r)=2$), and from the latter by either a sharp
        discontinuous transition with bistability for large values of
        $r$ or by a smooth transition for small $r$'s (Fig.4A).
        Observe that fluctuations, stemming from network
          sparsity, have blurred away the line of mean-field Hopf
        bifurcations as well as the BT point, so that the
        resulting intermediate phase is reminiscent of the mean-field
        excitable phase but, crucially, with a non-vanishing
          irregular activity (see below). 

        Importantly, even if the phase diagram in Fig.4A has been
        derived for annealed networks, qualitatively identical ones
        ---albeit with shifted phase boundaries--- can be
        computationally obtained for sparse networks with a fixed
        (quenched) architecture (such as 2D lattices and random
        regular networks) with the same values of $K$ and $\alpha$.
        Hence, the forthcoming results are, in general, valid for all
        these types of networks.

        First of all, we notice that the intermediate phase in Fig.4
        exhibits all the key features of cortical asynchronous states
        \cite{Renart,Doiron,Deneve,Hansel-Mato,Einevoll}, so we call
        it \emph{asynchronous (AS) phase}. In particular:
        ({\textbf{i}}) The coefficient of variation ($CV$)
        ---i.e. the ratio of the variance to the mean activity
          (see SI)--- takes values $CV>1$,
          as corresponds to highly irregular single-node activations
        (Fig.4D).  ({\textbf{ii}}) Time series for inhibitory nodes
        tightly follow excitatory ones (Fig.4E), leading to strong
        lagged cross-correlations between excitation and inhibition
        (Fig.4F/G), a feature absent in the standard active phase
          (Fig.4H/I). ({\textbf{iii}}) Small averaged pairwise
        correlations are found (not shown). However, most remarkably,
        the elucidated AS phase ---in the regime of large $\lambda$
        and $r$ values--- exhibits also important features
        characteristic of brain spontaneous activity  that are
          typically not described by standard simple models of
        asynchronous-states \cite{Renart}. These include: \textbf{(A)
          Collective excitability:} As shown in Fig.4B, the AS phase
        is characterized by a large degree of non-normality ---as
        quantified, e.g., by the Henrici index \cite{Lambiotte}---
        that grows with both $\lambda$ and $r$. In this regime, the AS
        phase can be highly excitable as illustrated in Fig.4J, so
        those small perturbations can give rise to very large
        excursions far away from quiescence, generating large tilted
        avalanches \textbf{(B) Bistability with hysteresis:} Given
        that the AS phase can coexist with the active one, it is
        feasible to shift the network dynamical regime from a
        low-activity (AS) to high-activity (standard active) one by
        perturbating the system above some threshold (see Fig.4J
        and SI). This shift resembles the striking empirical
        observation that the collective state of the cortex can be
        shifted from a low-activity state to a stable active state
        with a relatively small perturbation \cite{Yang-Dan}. Another
        consequence of bistability is the presence of hysteresis which
        is important for, e.g., working memory \cite{WM}.  \textbf{(C)
          Partial synchronization:} As illustrated in the SI 
          (Fig.S5), there are quasi-oscillations, evinced as a peak
        in the Fourier transform of the activity time series
        \cite{Wallace,Hidalgo}, followed by a power-law decay
        revealing variable and transient levels of synchronization, as
        observed in the cortex \cite{Wallace}.  Importantly, this
        phenomenology survives when inhibition to inhibitory neurons
        is switched on ($r_i \neq 0$), but tends to disappear as the
        symmetric limit ($r_i = r_e$) is approached (see SI),
        providing a simple explanation of why ``inhibition of
        inhibition'' is often mild in brain networks, which are thus
        ``inhibition dominated'' \cite{Scanziani}
        (cf.\cite{Sejnowski}).

        Finally, we also confirmed computationally that the phase
      transition from the quiescent to the AS phase is described by
        the directed-percolation class (both in mean-field and in
        2D; see SI).  Violations of such universality occurring at special
      points will be described elsewhere \cite{Helena}.

         In summary, the EI-CP ---an extension of the archetypical
         contact process including additionally inhibitory nodes--- 
      exhibits an extremely-rich phenomenology,
         especially in the inhibition-dominated case and on
           sparse networks.  In particular, on these networks, one
       finds an AS phase that captures the basic features of
       asynchronous states in the brain, and also describes additional
       remarkable properties, such as collective excitability and
       partial synchronization, which are usually not explained by
       existing simple models of asynchronous states.  In this
         way, the model allowed us to rationalize empirical
       observations such as (i) scale-free tilted neuronal avalanches
       \cite{zebra,Gleeson}, (ii) regime shifts in the overall network
       state emerging after a limited perturbation \cite{Yang-Dan},
       and (iii) quasi-oscillations \cite{Wallace}, that are certainly
       well-beyond the limit of validity of the standard BP picture,
       as well as simple models of asynchronous states \cite{Renart}.
       Furthermore, this allows us to put under the same
       parsimonious setting critical states (including some exotic
       ones) and asynchronous states, paving the way towards a deeper
       understanding of the statistical mechanics of spontaneous brain
       activity.  Extensions of our 
       approach, including important features of actual neural
         networks,
		essential for memory and learning,
       such as more heterogeneous network architectures,
       distributed synaptic-weights, refractory periods, etc.
       will be explored elsewhere.

\vspace{0.25cm} We acknowledge the Spanish Ministry and Agencia
Estatal de investigaci{\'o}n (AEI) through Project of I+D+i
Ref. PID2020-113681GB-I00, financed by MICIN/AEI/10.13039/501100011033
and FEDER “A way to make Europe”, as well as the Consejer{\'\i}a de
Conocimiento, Investigaci{\'o}n Universidad, Junta de Andaluc{\'\i}a
and European Regional Development Fund, Project references
A-FQM-175-UGR18 and P20-00173 for financial
support. R.C.L. acknoledges funding from the Spanish Ministry and AEI,
grant FPU19/03887. V.B. was supported by a Sofja Kovalevskaja Award
from the A. von Humboldt Foundation (German
Federal Ministry of Education and Research).  We also thank
H.C. Piuvezam, J. Pretel, G.B. Morales, P. Moretti, O. Vinogradov and
E. Giannakakis for valuable discussions.
 
\bibliographystyle{apsrev4-2}
%\bibliography{Bib-EICP-PRL.bib}

\begin{thebibliography}{77}%
\makeatletter
\providecommand \@ifxundefined [1]{%
 \@ifx{#1\undefined}
}%
\providecommand \@ifnum [1]{%
 \ifnum #1\expandafter \@firstoftwo
 \else \expandafter \@secondoftwo
 \fi
}%
\providecommand \@ifx [1]{%
 \ifx #1\expandafter \@firstoftwo
 \else \expandafter \@secondoftwo
 \fi
}%
\providecommand \natexlab [1]{#1}%
\providecommand \enquote  [1]{``#1''}%
\providecommand \bibnamefont  [1]{#1}%
\providecommand \bibfnamefont [1]{#1}%
\providecommand \citenamefont [1]{#1}%
\providecommand \href@noop [0]{\@secondoftwo}%
\providecommand \href [0]{\begingroup \@sanitize@url \@href}%
\providecommand \@href[1]{\@@startlink{#1}\@@href}%
\providecommand \@@href[1]{\endgroup#1\@@endlink}%
\providecommand \@sanitize@url [0]{\catcode `\\12\catcode `\$12\catcode
  `\&12\catcode `\#12\catcode `\^12\catcode `\_12\catcode `\%12\relax}%
\providecommand \@@startlink[1]{}%
\providecommand \@@endlink[0]{}%
\providecommand \url  [0]{\begingroup\@sanitize@url \@url }%
\providecommand \@url [1]{\endgroup\@href {#1}{\urlprefix }}%
\providecommand \urlprefix  [0]{URL }%
\providecommand \Eprint [0]{\href }%
\providecommand \doibase [0]{https://doi.org/}%
\providecommand \selectlanguage [0]{\@gobble}%
\providecommand \bibinfo  [0]{\@secondoftwo}%
\providecommand \bibfield  [0]{\@secondoftwo}%
\providecommand \translation [1]{[#1]}%
\providecommand \BibitemOpen [0]{}%
\providecommand \bibitemStop [0]{}%
\providecommand \bibitemNoStop [0]{.\EOS\space}%
\providecommand \EOS [0]{\spacefactor3000\relax}%
\providecommand \BibitemShut  [1]{\csname bibitem#1\endcsname}%
\let\auto@bib@innerbib\@empty
%</preamble>
\bibitem [{\citenamefont {Mora}\ and\ \citenamefont {Bialek}(2011)}]{Mora}%
  \BibitemOpen
  \bibfield  {author} {\bibinfo {author} {\bibfnamefont {T.}~\bibnamefont
  {Mora}}\ and\ \bibinfo {author} {\bibfnamefont {W.}~\bibnamefont {Bialek}},\
  }\href@noop {} {\bibfield  {journal} {\bibinfo  {journal} {J. Stat. Phys.}\
  }\textbf {\bibinfo {volume} {144}},\ \bibinfo {pages} {268} (\bibinfo {year}
  {2011})}\BibitemShut {NoStop}%
\bibitem [{\citenamefont {Mu{\~n}oz}(2018)}]{RMP}%
  \BibitemOpen
  \bibfield  {author} {\bibinfo {author} {\bibfnamefont {M.~A.}\ \bibnamefont
  {Mu{\~n}oz}},\ }\href@noop {} {\bibfield  {journal} {\bibinfo  {journal}
  {Rev. Mod. Phys.}\ }\textbf {\bibinfo {volume} {90}},\ \bibinfo {pages}
  {031001} (\bibinfo {year} {2018})}\BibitemShut {NoStop}%
\bibitem [{\citenamefont {Beggs}\ and\ \citenamefont {Plenz}(2003)}]{BP2003}%
  \BibitemOpen
  \bibfield  {author} {\bibinfo {author} {\bibfnamefont {J.~M.}\ \bibnamefont
  {Beggs}}\ and\ \bibinfo {author} {\bibfnamefont {D.}~\bibnamefont {Plenz}},\
  }\href@noop {} {\bibfield  {journal} {\bibinfo  {journal} {J. Neurosci.}\
  }\textbf {\bibinfo {volume} {23}},\ \bibinfo {pages} {11167} (\bibinfo {year}
  {2003})}\BibitemShut {NoStop}%
\bibitem [{\citenamefont {Petermann}\ \emph {et~al.}(2009)\citenamefont
  {Petermann}, \citenamefont {Thiagarajan}, \citenamefont {Lebedev},
  \citenamefont {Nicolelis}, \citenamefont {Chialvo},\ and\ \citenamefont
  {Plenz}}]{Petermann09}%
  \BibitemOpen
  \bibfield  {author} {\bibinfo {author} {\bibfnamefont {T.}~\bibnamefont
  {Petermann}}, \bibinfo {author} {\bibfnamefont {T.~C.}\ \bibnamefont
  {Thiagarajan}}, \bibinfo {author} {\bibfnamefont {M.~A.}\ \bibnamefont
  {Lebedev}}, \bibinfo {author} {\bibfnamefont {M.~A.}\ \bibnamefont
  {Nicolelis}}, \bibinfo {author} {\bibfnamefont {D.~R.}\ \bibnamefont
  {Chialvo}},\ and\ \bibinfo {author} {\bibfnamefont {D.}~\bibnamefont
  {Plenz}},\ }\href@noop {} {\bibfield  {journal} {\bibinfo  {journal}
  {Proceedings of the National Academy of Sciences}\ }\textbf {\bibinfo
  {volume} {106}},\ \bibinfo {pages} {15921} (\bibinfo {year}
  {2009})}\BibitemShut {NoStop}%
\bibitem [{\citenamefont {Haimovici}\ \emph {et~al.}(2013)\citenamefont
  {Haimovici}, \citenamefont {Tagliazucchi}, \citenamefont {Balenzuela},\ and\
  \citenamefont {Chialvo}}]{Haimovici}%
  \BibitemOpen
  \bibfield  {author} {\bibinfo {author} {\bibfnamefont {A.}~\bibnamefont
  {Haimovici}}, \bibinfo {author} {\bibfnamefont {E.}~\bibnamefont
  {Tagliazucchi}}, \bibinfo {author} {\bibfnamefont {P.}~\bibnamefont
  {Balenzuela}},\ and\ \bibinfo {author} {\bibfnamefont {D.~R.}\ \bibnamefont
  {Chialvo}},\ }\href@noop {} {\bibfield  {journal} {\bibinfo  {journal} {Phys.
  Rev. Lett.}\ }\textbf {\bibinfo {volume} {110}},\ \bibinfo {pages} {178101}
  (\bibinfo {year} {2013})}\BibitemShut {NoStop}%
\bibitem [{\citenamefont {Tagliazucchi}\ \emph {et~al.}(2012)\citenamefont
  {Tagliazucchi}, \citenamefont {Balenzuela}, \citenamefont {Fraiman},\ and\
  \citenamefont {Chialvo}}]{Taglia}%
  \BibitemOpen
  \bibfield  {author} {\bibinfo {author} {\bibfnamefont {E.}~\bibnamefont
  {Tagliazucchi}}, \bibinfo {author} {\bibfnamefont {P.}~\bibnamefont
  {Balenzuela}}, \bibinfo {author} {\bibfnamefont {D.}~\bibnamefont
  {Fraiman}},\ and\ \bibinfo {author} {\bibfnamefont {D.~R.}\ \bibnamefont
  {Chialvo}},\ }\href@noop {} {\bibfield  {journal} {\bibinfo  {journal}
  {Front. Physiol.}\ }\textbf {\bibinfo {volume} {3}},\ \bibinfo {pages} {15}
  (\bibinfo {year} {2012})}\BibitemShut {NoStop}%
\bibitem [{\citenamefont {Shriki}\ \emph {et~al.}(2013)\citenamefont {Shriki},
  \citenamefont {Alstott}, \citenamefont {Carver}, \citenamefont {Holroyd},
  \citenamefont {Henson}, \citenamefont {Smith}, \citenamefont {Coppola},
  \citenamefont {Bullmore},\ and\ \citenamefont {Plenz}}]{Plenz-Shriki2013}%
  \BibitemOpen
  \bibfield  {author} {\bibinfo {author} {\bibfnamefont {O.}~\bibnamefont
  {Shriki}}, \bibinfo {author} {\bibfnamefont {J.}~\bibnamefont {Alstott}},
  \bibinfo {author} {\bibfnamefont {F.}~\bibnamefont {Carver}}, \bibinfo
  {author} {\bibfnamefont {T.}~\bibnamefont {Holroyd}}, \bibinfo {author}
  {\bibfnamefont {R.~N.}\ \bibnamefont {Henson}}, \bibinfo {author}
  {\bibfnamefont {M.~L.}\ \bibnamefont {Smith}}, \bibinfo {author}
  {\bibfnamefont {R.}~\bibnamefont {Coppola}}, \bibinfo {author} {\bibfnamefont
  {E.}~\bibnamefont {Bullmore}},\ and\ \bibinfo {author} {\bibfnamefont
  {D.}~\bibnamefont {Plenz}},\ }\href@noop {} {\bibfield  {journal} {\bibinfo
  {journal} {J. Neurosci.}\ }\textbf {\bibinfo {volume} {33}},\ \bibinfo
  {pages} {7079} (\bibinfo {year} {2013})}\BibitemShut {NoStop}%
\bibitem [{\citenamefont {Bellay}\ \emph {et~al.}(2015)\citenamefont {Bellay},
  \citenamefont {Klaus}, \citenamefont {Seshadri},\ and\ \citenamefont
  {Plenz}}]{2-photon}%
  \BibitemOpen
  \bibfield  {author} {\bibinfo {author} {\bibfnamefont {T.}~\bibnamefont
  {Bellay}}, \bibinfo {author} {\bibfnamefont {A.}~\bibnamefont {Klaus}},
  \bibinfo {author} {\bibfnamefont {S.}~\bibnamefont {Seshadri}},\ and\
  \bibinfo {author} {\bibfnamefont {D.}~\bibnamefont {Plenz}},\ }\href@noop {}
  {\bibfield  {journal} {\bibinfo  {journal} {Elife}\ }\textbf {\bibinfo
  {volume} {4}},\ \bibinfo {pages} {e07224} (\bibinfo {year}
  {2015})}\BibitemShut {NoStop}%
\bibitem [{\citenamefont {Yu}\ \emph {et~al.}(2017)\citenamefont {Yu},
  \citenamefont {Ribeiro}, \citenamefont {Meisel}, \citenamefont {Chou},
  \citenamefont {Mitz}, \citenamefont {Saunders},\ and\ \citenamefont
  {Plenz}}]{Plenz-task}%
  \BibitemOpen
  \bibfield  {author} {\bibinfo {author} {\bibfnamefont {S.}~\bibnamefont
  {Yu}}, \bibinfo {author} {\bibfnamefont {T.~L.}\ \bibnamefont {Ribeiro}},
  \bibinfo {author} {\bibfnamefont {C.}~\bibnamefont {Meisel}}, \bibinfo
  {author} {\bibfnamefont {S.}~\bibnamefont {Chou}}, \bibinfo {author}
  {\bibfnamefont {A.}~\bibnamefont {Mitz}}, \bibinfo {author} {\bibfnamefont
  {R.}~\bibnamefont {Saunders}},\ and\ \bibinfo {author} {\bibfnamefont
  {D.}~\bibnamefont {Plenz}},\ }\href@noop {} {\bibfield  {journal} {\bibinfo
  {journal} {Elife}\ }\textbf {\bibinfo {volume} {6}},\ \bibinfo {pages}
  {e27119} (\bibinfo {year} {2017})}\BibitemShut {NoStop}%
\bibitem [{\citenamefont {Liggett}(2004)}]{Liggett}%
  \BibitemOpen
  \bibfield  {author} {\bibinfo {author} {\bibfnamefont {T.}~\bibnamefont
  {Liggett}},\ }\href {http://books.google.it/books?id=I3aNPR1FursC} {\emph
  {\bibinfo {title} {Interacting Particle Systems}}},\ Classics in Mathematics\
  (\bibinfo  {publisher} {Springer},\ \bibinfo {year} {2004})\BibitemShut
  {NoStop}%
\bibitem [{\citenamefont {Harris}(2002)}]{Harris}%
  \BibitemOpen
  \bibfield  {author} {\bibinfo {author} {\bibfnamefont {T.~E.}\ \bibnamefont
  {Harris}},\ }\href@noop {} {\emph {\bibinfo {title} {The theory of branching
  processes}}}\ (\bibinfo  {publisher} {Courier Corporation},\ \bibinfo {year}
  {2002})\BibitemShut {NoStop}%
\bibitem [{\citenamefont {Sethna}\ \emph {et~al.}(2001)\citenamefont {Sethna},
  \citenamefont {Dahmen},\ and\ \citenamefont {Myers}}]{Sethna}%
  \BibitemOpen
  \bibfield  {author} {\bibinfo {author} {\bibfnamefont {J.~P.}\ \bibnamefont
  {Sethna}}, \bibinfo {author} {\bibfnamefont {K.~A.}\ \bibnamefont {Dahmen}},\
  and\ \bibinfo {author} {\bibfnamefont {C.~R.}\ \bibnamefont {Myers}},\
  }\href@noop {} {\bibfield  {journal} {\bibinfo  {journal} {Nature}\ }\textbf
  {\bibinfo {volume} {410}},\ \bibinfo {pages} {242} (\bibinfo {year}
  {2001})}\BibitemShut {NoStop}%
\bibitem [{\citenamefont {Friedman}\ \emph {et~al.}(2012)\citenamefont
  {Friedman}, \citenamefont {Ito}, \citenamefont {Brinkman}, \citenamefont
  {Shimono}, \citenamefont {DeVille}, \citenamefont {Dahmen}, \citenamefont
  {Beggs},\ and\ \citenamefont {Butler}}]{shape}%
  \BibitemOpen
  \bibfield  {author} {\bibinfo {author} {\bibfnamefont {N.}~\bibnamefont
  {Friedman}}, \bibinfo {author} {\bibfnamefont {S.}~\bibnamefont {Ito}},
  \bibinfo {author} {\bibfnamefont {B.~A.~W.}\ \bibnamefont {Brinkman}},
  \bibinfo {author} {\bibfnamefont {M.}~\bibnamefont {Shimono}}, \bibinfo
  {author} {\bibfnamefont {R.~E.~L.}\ \bibnamefont {DeVille}}, \bibinfo
  {author} {\bibfnamefont {K.~A.}\ \bibnamefont {Dahmen}}, \bibinfo {author}
  {\bibfnamefont {J.~M.}\ \bibnamefont {Beggs}},\ and\ \bibinfo {author}
  {\bibfnamefont {T.~C.}\ \bibnamefont {Butler}},\ }\href
  {https://doi.org/10.1103/PhysRevLett.108.208102} {\bibfield  {journal}
  {\bibinfo  {journal} {Phys. Rev. Lett.}\ }\textbf {\bibinfo {volume} {108}},\
  \bibinfo {pages} {208102} (\bibinfo {year} {2012})}\BibitemShut {NoStop}%
\bibitem [{\citenamefont {Miller}\ \emph {et~al.}(2021)\citenamefont {Miller},
  \citenamefont {Yu}, \citenamefont {Pajevic},\ and\ \citenamefont
  {Plenz}}]{Plenz2021}%
  \BibitemOpen
  \bibfield  {author} {\bibinfo {author} {\bibfnamefont {S.~R.}\ \bibnamefont
  {Miller}}, \bibinfo {author} {\bibfnamefont {S.}~\bibnamefont {Yu}}, \bibinfo
  {author} {\bibfnamefont {S.}~\bibnamefont {Pajevic}},\ and\ \bibinfo {author}
  {\bibfnamefont {D.}~\bibnamefont {Plenz}},\ }\href@noop {} {\bibfield
  {journal} {\bibinfo  {journal} {Network Neuroscience}\ }\textbf {\bibinfo
  {volume} {5}},\ \bibinfo {pages} {505} (\bibinfo {year} {2021})}\BibitemShut
  {NoStop}%
\bibitem [{\citenamefont {di~Santo}\ \emph {et~al.}(2017)\citenamefont
  {di~Santo}, \citenamefont {Villegas}, \citenamefont {Burioni},\ and\
  \citenamefont {Mu{\~n}oz}}]{Serena-BP}%
  \BibitemOpen
  \bibfield  {author} {\bibinfo {author} {\bibfnamefont {S.}~\bibnamefont
  {di~Santo}}, \bibinfo {author} {\bibfnamefont {P.}~\bibnamefont {Villegas}},
  \bibinfo {author} {\bibfnamefont {R.}~\bibnamefont {Burioni}},\ and\ \bibinfo
  {author} {\bibfnamefont {M.~A.}\ \bibnamefont {Mu{\~n}oz}},\ }\href@noop {}
  {\bibfield  {journal} {\bibinfo  {journal} {Phys. Rev. E}\ }\textbf {\bibinfo
  {volume} {95}},\ \bibinfo {pages} {032115} (\bibinfo {year}
  {2017})}\BibitemShut {NoStop}%
\bibitem [{\citenamefont {Priesemann}\ \emph {et~al.}(2009)\citenamefont
  {Priesemann}, \citenamefont {Munk},\ and\ \citenamefont {Wibral}}]{Viola1}%
  \BibitemOpen
  \bibfield  {author} {\bibinfo {author} {\bibfnamefont {V.}~\bibnamefont
  {Priesemann}}, \bibinfo {author} {\bibfnamefont {M.~H.}\ \bibnamefont
  {Munk}},\ and\ \bibinfo {author} {\bibfnamefont {M.}~\bibnamefont {Wibral}},\
  }\href@noop {} {\bibfield  {journal} {\bibinfo  {journal} {BMC neuroscience}\
  }\textbf {\bibinfo {volume} {10}},\ \bibinfo {pages} {1} (\bibinfo {year}
  {2009})}\BibitemShut {NoStop}%
\bibitem [{\citenamefont {Martinello}\ \emph {et~al.}(2017)\citenamefont
  {Martinello}, \citenamefont {Hidalgo}, \citenamefont {Maritan}, \citenamefont
  {di~Santo}, \citenamefont {Plenz},\ and\ \citenamefont
  {Mu{\~n}oz}}]{neutral}%
  \BibitemOpen
  \bibfield  {author} {\bibinfo {author} {\bibfnamefont {M.}~\bibnamefont
  {Martinello}}, \bibinfo {author} {\bibfnamefont {J.}~\bibnamefont {Hidalgo}},
  \bibinfo {author} {\bibfnamefont {A.}~\bibnamefont {Maritan}}, \bibinfo
  {author} {\bibfnamefont {S.}~\bibnamefont {di~Santo}}, \bibinfo {author}
  {\bibfnamefont {D.}~\bibnamefont {Plenz}},\ and\ \bibinfo {author}
  {\bibfnamefont {M.~A.}\ \bibnamefont {Mu{\~n}oz}},\ }\href@noop {} {\bibfield
   {journal} {\bibinfo  {journal} {Physical Review X}\ }\textbf {\bibinfo
  {volume} {7}},\ \bibinfo {pages} {041071} (\bibinfo {year}
  {2017})}\BibitemShut {NoStop}%
\bibitem [{\citenamefont {Priesemann}\ \emph {et~al.}(2013)\citenamefont
  {Priesemann}, \citenamefont {Valderrama}, \citenamefont {Wibral},\ and\
  \citenamefont {Le~Van~Quyen}}]{Viola2}%
  \BibitemOpen
  \bibfield  {author} {\bibinfo {author} {\bibfnamefont {V.}~\bibnamefont
  {Priesemann}}, \bibinfo {author} {\bibfnamefont {M.}~\bibnamefont
  {Valderrama}}, \bibinfo {author} {\bibfnamefont {M.}~\bibnamefont {Wibral}},\
  and\ \bibinfo {author} {\bibfnamefont {M.}~\bibnamefont {Le~Van~Quyen}},\
  }\href@noop {} {\bibfield  {journal} {\bibinfo  {journal} {PLoS computational
  biology}\ }\textbf {\bibinfo {volume} {9}},\ \bibinfo {pages} {e1002985}
  (\bibinfo {year} {2013})}\BibitemShut {NoStop}%
\bibitem [{\citenamefont {Fontenele}\ \emph {et~al.}(2019)\citenamefont
  {Fontenele}, \citenamefont {de~Vasconcelos}, \citenamefont {Feliciano},
  \citenamefont {Aguiar}, \citenamefont {Soares-Cunha}, \citenamefont
  {Coimbra}, \citenamefont {Dalla~Porta}, \citenamefont {Ribeiro},
  \citenamefont {Rodrigues}, \citenamefont {Sousa} \emph {et~al.}}]{Fontenele}%
  \BibitemOpen
  \bibfield  {author} {\bibinfo {author} {\bibfnamefont {A.~J.}\ \bibnamefont
  {Fontenele}}, \bibinfo {author} {\bibfnamefont {N.~A.}\ \bibnamefont
  {de~Vasconcelos}}, \bibinfo {author} {\bibfnamefont {T.}~\bibnamefont
  {Feliciano}}, \bibinfo {author} {\bibfnamefont {L.~A.}\ \bibnamefont
  {Aguiar}}, \bibinfo {author} {\bibfnamefont {C.}~\bibnamefont
  {Soares-Cunha}}, \bibinfo {author} {\bibfnamefont {B.}~\bibnamefont
  {Coimbra}}, \bibinfo {author} {\bibfnamefont {L.}~\bibnamefont
  {Dalla~Porta}}, \bibinfo {author} {\bibfnamefont {S.}~\bibnamefont
  {Ribeiro}}, \bibinfo {author} {\bibfnamefont {A.~J.}\ \bibnamefont
  {Rodrigues}}, \bibinfo {author} {\bibfnamefont {N.}~\bibnamefont {Sousa}},
  \emph {et~al.},\ }\href@noop {} {\bibfield  {journal} {\bibinfo  {journal}
  {Physical Review Letters}\ }\textbf {\bibinfo {volume} {122}},\ \bibinfo
  {pages} {208101} (\bibinfo {year} {2019})}\BibitemShut {NoStop}%
\bibitem [{\citenamefont {Ponce-Alvarez}\ \emph
  {et~al.}(2018{\natexlab{a}})\citenamefont {Ponce-Alvarez}, \citenamefont
  {Jouary}, \citenamefont {Privat}, \citenamefont {Deco},\ and\ \citenamefont
  {Sumbre}}]{Crackling}%
  \BibitemOpen
  \bibfield  {author} {\bibinfo {author} {\bibfnamefont {A.}~\bibnamefont
  {Ponce-Alvarez}}, \bibinfo {author} {\bibfnamefont {A.}~\bibnamefont
  {Jouary}}, \bibinfo {author} {\bibfnamefont {M.}~\bibnamefont {Privat}},
  \bibinfo {author} {\bibfnamefont {G.}~\bibnamefont {Deco}},\ and\ \bibinfo
  {author} {\bibfnamefont {G.}~\bibnamefont {Sumbre}},\ }\href@noop {}
  {\bibfield  {journal} {\bibinfo  {journal} {Neuron}\ }\textbf {\bibinfo
  {volume} {100}},\ \bibinfo {pages} {1446} (\bibinfo {year}
  {2018}{\natexlab{a}})}\BibitemShut {NoStop}%
\bibitem [{\citenamefont {Di~Santo}\ \emph {et~al.}(2018)\citenamefont
  {Di~Santo}, \citenamefont {Villegas}, \citenamefont {Burioni},\ and\
  \citenamefont {Mu{\~n}oz}}]{LG}%
  \BibitemOpen
  \bibfield  {author} {\bibinfo {author} {\bibfnamefont {S.}~\bibnamefont
  {Di~Santo}}, \bibinfo {author} {\bibfnamefont {P.}~\bibnamefont {Villegas}},
  \bibinfo {author} {\bibfnamefont {R.}~\bibnamefont {Burioni}},\ and\ \bibinfo
  {author} {\bibfnamefont {M.~A.}\ \bibnamefont {Mu{\~n}oz}},\ }\href@noop {}
  {\bibfield  {journal} {\bibinfo  {journal} {Proceedings of the National
  Academy of Sciences}\ }\textbf {\bibinfo {volume} {115}},\ \bibinfo {pages}
  {E1356} (\bibinfo {year} {2018})}\BibitemShut {NoStop}%
\bibitem [{\citenamefont {Buend{\'\i}a}\ \emph {et~al.}(2021)\citenamefont
  {Buend{\'\i}a}, \citenamefont {Villegas}, \citenamefont {Burioni},\ and\
  \citenamefont {Mu{\~n}oz}}]{hybrid}%
  \BibitemOpen
  \bibfield  {author} {\bibinfo {author} {\bibfnamefont {V.}~\bibnamefont
  {Buend{\'\i}a}}, \bibinfo {author} {\bibfnamefont {P.}~\bibnamefont
  {Villegas}}, \bibinfo {author} {\bibfnamefont {R.}~\bibnamefont {Burioni}},\
  and\ \bibinfo {author} {\bibfnamefont {M.~A.}\ \bibnamefont {Mu{\~n}oz}},\
  }\href@noop {} {\bibfield  {journal} {\bibinfo  {journal} {Physical Review
  Research}\ }\textbf {\bibinfo {volume} {3}},\ \bibinfo {pages} {023224}
  (\bibinfo {year} {2021})}\BibitemShut {NoStop}%
\bibitem [{\citenamefont {Mariani}\ \emph {et~al.}(2022)\citenamefont
  {Mariani}, \citenamefont {Nicoletti}, \citenamefont {Bisio}, \citenamefont
  {Maschietto}, \citenamefont {Vassanelli},\ and\ \citenamefont
  {Suweis}}]{Suweis}%
  \BibitemOpen
  \bibfield  {author} {\bibinfo {author} {\bibfnamefont {B.}~\bibnamefont
  {Mariani}}, \bibinfo {author} {\bibfnamefont {G.}~\bibnamefont {Nicoletti}},
  \bibinfo {author} {\bibfnamefont {M.}~\bibnamefont {Bisio}}, \bibinfo
  {author} {\bibfnamefont {M.}~\bibnamefont {Maschietto}}, \bibinfo {author}
  {\bibfnamefont {S.}~\bibnamefont {Vassanelli}},\ and\ \bibinfo {author}
  {\bibfnamefont {S.}~\bibnamefont {Suweis}},\ }\href@noop {} {\bibfield
  {journal} {\bibinfo  {journal} {Scientific reports}\ }\textbf {\bibinfo
  {volume} {12}},\ \bibinfo {pages} {1} (\bibinfo {year} {2022})}\BibitemShut
  {NoStop}%
\bibitem [{\citenamefont {Plenz}\ and\ \citenamefont
  {Niebur}(2014)}]{Schuster}%
  \BibitemOpen
  \bibfield  {author} {\bibinfo {author} {\bibfnamefont {D.}~\bibnamefont
  {Plenz}}\ and\ \bibinfo {author} {\bibfnamefont {E.}~\bibnamefont {Niebur}},\
  }\href@noop {} {\emph {\bibinfo {title} {Criticality in neural systems}}}\
  (\bibinfo  {publisher} {John Wiley \& Sons},\ \bibinfo {year}
  {2014})\BibitemShut {NoStop}%
\bibitem [{\citenamefont {Chialvo}(2010)}]{Chialvo2010}%
  \BibitemOpen
  \bibfield  {author} {\bibinfo {author} {\bibfnamefont {D.~R.}\ \bibnamefont
  {Chialvo}},\ }\href@noop {} {\bibfield  {journal} {\bibinfo  {journal} {Nat.
  Phys.}\ }\textbf {\bibinfo {volume} {6}},\ \bibinfo {pages} {744} (\bibinfo
  {year} {2010})}\BibitemShut {NoStop}%
\bibitem [{\citenamefont {Levina}\ \emph {et~al.}(2007)\citenamefont {Levina},
  \citenamefont {Herrmann},\ and\ \citenamefont {Geisel}}]{Levina}%
  \BibitemOpen
  \bibfield  {author} {\bibinfo {author} {\bibfnamefont {A.}~\bibnamefont
  {Levina}}, \bibinfo {author} {\bibfnamefont {J.~M.}\ \bibnamefont
  {Herrmann}},\ and\ \bibinfo {author} {\bibfnamefont {T.}~\bibnamefont
  {Geisel}},\ }\href@noop {} {\bibfield  {journal} {\bibinfo  {journal} {Nature
  physics}\ }\textbf {\bibinfo {volume} {3}},\ \bibinfo {pages} {857} (\bibinfo
  {year} {2007})}\BibitemShut {NoStop}%
\bibitem [{\citenamefont {Fosque}\ \emph {et~al.}(2021)\citenamefont {Fosque},
  \citenamefont {Williams-Garc\'{\i}a}, \citenamefont {Beggs},\ and\
  \citenamefont {Ortiz}}]{Beggs2021}%
  \BibitemOpen
  \bibfield  {author} {\bibinfo {author} {\bibfnamefont {L.~J.}\ \bibnamefont
  {Fosque}}, \bibinfo {author} {\bibfnamefont {R.~V.}\ \bibnamefont
  {Williams-Garc\'{\i}a}}, \bibinfo {author} {\bibfnamefont {J.~M.}\
  \bibnamefont {Beggs}},\ and\ \bibinfo {author} {\bibfnamefont
  {G.}~\bibnamefont {Ortiz}},\ }\href
  {https://doi.org/10.1103/PhysRevLett.126.098101} {\bibfield  {journal}
  {\bibinfo  {journal} {Phys. Rev. Lett.}\ }\textbf {\bibinfo {volume} {126}},\
  \bibinfo {pages} {098101} (\bibinfo {year} {2021})}\BibitemShut {NoStop}%
\bibitem [{\citenamefont {Kinouchi}\ and\ \citenamefont
  {Copelli}(2006)}]{Kinouchi}%
  \BibitemOpen
  \bibfield  {author} {\bibinfo {author} {\bibfnamefont {O.}~\bibnamefont
  {Kinouchi}}\ and\ \bibinfo {author} {\bibfnamefont {M.}~\bibnamefont
  {Copelli}},\ }\href {https://doi.org/10.1038/nphys289} {\bibfield  {journal}
  {\bibinfo  {journal} {Nature Physics}\ }\textbf {\bibinfo {volume} {2}},\
  \bibinfo {pages} {348} (\bibinfo {year} {2006})}\BibitemShut {NoStop}%
\bibitem [{\citenamefont {Shew}\ \emph {et~al.}()\citenamefont {Shew},
  \citenamefont {Yang}, \citenamefont {Petermann}, \citenamefont {Roy},\ and\
  \citenamefont {Plenz}}]{shew_optimal}%
  \BibitemOpen
  \bibfield  {author} {\bibinfo {author} {\bibfnamefont {W.~L.}\ \bibnamefont
  {Shew}}, \bibinfo {author} {\bibfnamefont {H.}~\bibnamefont {Yang}}, \bibinfo
  {author} {\bibfnamefont {T.}~\bibnamefont {Petermann}}, \bibinfo {author}
  {\bibfnamefont {R.}~\bibnamefont {Roy}},\ and\ \bibinfo {author}
  {\bibfnamefont {D.}~\bibnamefont {Plenz}},\ }\href
  {https://doi.org/10.1523/JNEUROSCI.3864-09.2009} {\bibfield  {journal}
  {\bibinfo  {journal} {Journal of Neuroscience}\ }\textbf {\bibinfo {volume}
  {29}},\ \bibinfo {pages} {15595}},\ \Eprint {https://arxiv.org/abs/20007483}
  {20007483} \BibitemShut {NoStop}%
\bibitem [{\citenamefont {Shew}\ and\ \citenamefont
  {Plenz}(2013)}]{Plenz-functional}%
  \BibitemOpen
  \bibfield  {author} {\bibinfo {author} {\bibfnamefont {W.~L.}\ \bibnamefont
  {Shew}}\ and\ \bibinfo {author} {\bibfnamefont {D.}~\bibnamefont {Plenz}},\
  }\href@noop {} {\bibfield  {journal} {\bibinfo  {journal} {The
  Neuroscientist}\ }\textbf {\bibinfo {volume} {19}},\ \bibinfo {pages} {88}
  (\bibinfo {year} {2013})}\BibitemShut {NoStop}%
\bibitem [{\citenamefont {Ma}\ \emph {et~al.}(2019)\citenamefont {Ma},
  \citenamefont {Turrigiano}, \citenamefont {Wessel},\ and\ \citenamefont
  {Hengen}}]{Turrigiano}%
  \BibitemOpen
  \bibfield  {author} {\bibinfo {author} {\bibfnamefont {Z.}~\bibnamefont
  {Ma}}, \bibinfo {author} {\bibfnamefont {G.~G.}\ \bibnamefont {Turrigiano}},
  \bibinfo {author} {\bibfnamefont {R.}~\bibnamefont {Wessel}},\ and\ \bibinfo
  {author} {\bibfnamefont {K.~B.}\ \bibnamefont {Hengen}},\ }\href@noop {}
  {\bibfield  {journal} {\bibinfo  {journal} {Neuron}\ }\textbf {\bibinfo
  {volume} {104}},\ \bibinfo {pages} {655} (\bibinfo {year}
  {2019})}\BibitemShut {NoStop}%
\bibitem [{\citenamefont {Plenz}\ \emph {et~al.}(2021)\citenamefont {Plenz},
  \citenamefont {Ribeiro}, \citenamefont {Miller}, \citenamefont {Kells},
  \citenamefont {Vakili},\ and\ \citenamefont {Capek}}]{Plenz-review}%
  \BibitemOpen
  \bibfield  {author} {\bibinfo {author} {\bibfnamefont {D.}~\bibnamefont
  {Plenz}}, \bibinfo {author} {\bibfnamefont {T.~L.}\ \bibnamefont {Ribeiro}},
  \bibinfo {author} {\bibfnamefont {S.~R.}\ \bibnamefont {Miller}}, \bibinfo
  {author} {\bibfnamefont {P.~A.}\ \bibnamefont {Kells}}, \bibinfo {author}
  {\bibfnamefont {A.}~\bibnamefont {Vakili}},\ and\ \bibinfo {author}
  {\bibfnamefont {E.~L.}\ \bibnamefont {Capek}},\ }\href@noop {} {\bibfield
  {journal} {\bibinfo  {journal} {arXiv preprint arXiv:2102.09124}\ } (\bibinfo
  {year} {2021})}\BibitemShut {NoStop}%
\bibitem [{\citenamefont {Cocchi}\ \emph {et~al.}(2017)\citenamefont {Cocchi},
  \citenamefont {Gollo}, \citenamefont {Zalesky},\ and\ \citenamefont
  {Breakspear}}]{Breakspear-review}%
  \BibitemOpen
  \bibfield  {author} {\bibinfo {author} {\bibfnamefont {L.}~\bibnamefont
  {Cocchi}}, \bibinfo {author} {\bibfnamefont {L.~L.}\ \bibnamefont {Gollo}},
  \bibinfo {author} {\bibfnamefont {A.}~\bibnamefont {Zalesky}},\ and\ \bibinfo
  {author} {\bibfnamefont {M.}~\bibnamefont {Breakspear}},\ }\href@noop {}
  {\bibfield  {journal} {\bibinfo  {journal} {Progress in Neurobiology}\ }
  (\bibinfo {year} {2017})}\BibitemShut {NoStop}%
\bibitem [{\citenamefont {Wilting}\ and\ \citenamefont
  {Priesemann}(2019)}]{Viola-review}%
  \BibitemOpen
  \bibfield  {author} {\bibinfo {author} {\bibfnamefont {J.}~\bibnamefont
  {Wilting}}\ and\ \bibinfo {author} {\bibfnamefont {V.}~\bibnamefont
  {Priesemann}},\ }\href@noop {} {\bibfield  {journal} {\bibinfo  {journal}
  {Current opinion in neurobiology}\ }\textbf {\bibinfo {volume} {58}},\
  \bibinfo {pages} {105} (\bibinfo {year} {2019})}\BibitemShut {NoStop}%
\bibitem [{\citenamefont {Dayan}\ and\ \citenamefont {Abbott}(2006)}]{Dayan}%
  \BibitemOpen
  \bibfield  {author} {\bibinfo {author} {\bibfnamefont {P.}~\bibnamefont
  {Dayan}}\ and\ \bibinfo {author} {\bibfnamefont {L.~F.}\ \bibnamefont
  {Abbott}},\ }\href@noop {} {\emph {\bibinfo {title} {{ Theoretical
  Neuroscience: Computational and Mathematical Modeling of Neural Systems}}}}\
  (\bibinfo  {publisher} {Cambridge, MIT Press, USA},\ \bibinfo {year}
  {2006})\BibitemShut {NoStop}%
\bibitem [{\citenamefont {Isaacson}\ and\ \citenamefont
  {Scanziani}(2011)}]{Scanziani}%
  \BibitemOpen
  \bibfield  {author} {\bibinfo {author} {\bibfnamefont {J.~S.}\ \bibnamefont
  {Isaacson}}\ and\ \bibinfo {author} {\bibfnamefont {M.}~\bibnamefont
  {Scanziani}},\ }\href@noop {} {\bibfield  {journal} {\bibinfo  {journal}
  {Neuron}\ }\textbf {\bibinfo {volume} {72}},\ \bibinfo {pages} {231}
  (\bibinfo {year} {2011})}\BibitemShut {NoStop}%
\bibitem [{\citenamefont {Doiron}\ \emph {et~al.}(2016)\citenamefont {Doiron},
  \citenamefont {Litwin-Kumar}, \citenamefont {Rosenbaum}, \citenamefont
  {Ocker},\ and\ \citenamefont {Josi{\'c}}}]{Doiron}%
  \BibitemOpen
  \bibfield  {author} {\bibinfo {author} {\bibfnamefont {B.}~\bibnamefont
  {Doiron}}, \bibinfo {author} {\bibfnamefont {A.}~\bibnamefont
  {Litwin-Kumar}}, \bibinfo {author} {\bibfnamefont {R.}~\bibnamefont
  {Rosenbaum}}, \bibinfo {author} {\bibfnamefont {G.~K.}\ \bibnamefont
  {Ocker}},\ and\ \bibinfo {author} {\bibfnamefont {K.}~\bibnamefont
  {Josi{\'c}}},\ }\href@noop {} {\bibfield  {journal} {\bibinfo  {journal}
  {Nature neuroscience}\ }\textbf {\bibinfo {volume} {19}},\ \bibinfo {pages}
  {383} (\bibinfo {year} {2016})}\BibitemShut {NoStop}%
\bibitem [{\citenamefont {Den\'eve}\ and\ \citenamefont
  {Machens}(2016)}]{Deneve}%
  \BibitemOpen
  \bibfield  {author} {\bibinfo {author} {\bibfnamefont {S.}~\bibnamefont
  {Den\'eve}}\ and\ \bibinfo {author} {\bibfnamefont {C.~K.}\ \bibnamefont
  {Machens}},\ }\href {https://doi.org/10.1038/nn.4243} {\bibfield  {journal}
  {\bibinfo  {journal} {Nature Neuroscience}\ }\textbf {\bibinfo {volume}
  {19}},\ \bibinfo {pages} {375} (\bibinfo {year} {2016})}\BibitemShut
  {NoStop}%
\bibitem [{\citenamefont {van Vreeswijk}\ and\ \citenamefont
  {Sompolinsky}(1996)}]{vVS}%
  \BibitemOpen
  \bibfield  {author} {\bibinfo {author} {\bibfnamefont {C.}~\bibnamefont {van
  Vreeswijk}}\ and\ \bibinfo {author} {\bibfnamefont {H.}~\bibnamefont
  {Sompolinsky}},\ }\href {https://doi.org/10.1126/science.274.5293.1724}
  {\bibfield  {journal} {\bibinfo  {journal} {Science}\ }\textbf {\bibinfo
  {volume} {274}},\ \bibinfo {pages} {1724} (\bibinfo {year} {1996})},\ \Eprint
  {https://arxiv.org/abs/8939866} {8939866} \BibitemShut {NoStop}%
\bibitem [{\citenamefont {Brunel}(2000)}]{Brunel}%
  \BibitemOpen
  \bibfield  {author} {\bibinfo {author} {\bibfnamefont {N.}~\bibnamefont
  {Brunel}},\ }\href@noop {} {\bibfield  {journal} {\bibinfo  {journal}
  {Journal of computational neuroscience}\ }\textbf {\bibinfo {volume} {8}},\
  \bibinfo {pages} {183} (\bibinfo {year} {2000})}\BibitemShut {NoStop}%
\bibitem [{\citenamefont {Brunel}\ and\ \citenamefont {Wang}(2003)}]{Brunel2}%
  \BibitemOpen
  \bibfield  {author} {\bibinfo {author} {\bibfnamefont {N.}~\bibnamefont
  {Brunel}}\ and\ \bibinfo {author} {\bibfnamefont {X.-J.}\ \bibnamefont
  {Wang}},\ }\href {https://doi.org/10.1152/jn.01095.2002} {\bibfield
  {journal} {\bibinfo  {journal} {Journal of Neurophysiology}\ }\textbf
  {\bibinfo {volume} {90}},\ \bibinfo {pages} {415} (\bibinfo {year} {2003})},\
  \bibinfo {note} {pMID: 12611969}\BibitemShut {NoStop}%
\bibitem [{\citenamefont {Ginzburg}\ and\ \citenamefont
  {Sompolinsky}(1994)}]{Sompo1}%
  \BibitemOpen
  \bibfield  {author} {\bibinfo {author} {\bibfnamefont {I.}~\bibnamefont
  {Ginzburg}}\ and\ \bibinfo {author} {\bibfnamefont {H.}~\bibnamefont
  {Sompolinsky}},\ }\href {https://doi.org/10.1103/PhysRevE.50.3171} {\bibfield
   {journal} {\bibinfo  {journal} {Phys. Rev. E}\ }\textbf {\bibinfo {volume}
  {50}},\ \bibinfo {pages} {3171} (\bibinfo {year} {1994})}\BibitemShut
  {NoStop}%
\bibitem [{\citenamefont {Renart}\ \emph {et~al.}(2010)\citenamefont {Renart},
  \citenamefont {De~La~Rocha}, \citenamefont {Bartho}, \citenamefont
  {Hollender}, \citenamefont {Parga}, \citenamefont {Reyes},\ and\
  \citenamefont {Harris}}]{Renart}%
  \BibitemOpen
  \bibfield  {author} {\bibinfo {author} {\bibfnamefont {A.}~\bibnamefont
  {Renart}}, \bibinfo {author} {\bibfnamefont {J.}~\bibnamefont {De~La~Rocha}},
  \bibinfo {author} {\bibfnamefont {P.}~\bibnamefont {Bartho}}, \bibinfo
  {author} {\bibfnamefont {L.}~\bibnamefont {Hollender}}, \bibinfo {author}
  {\bibfnamefont {N.}~\bibnamefont {Parga}}, \bibinfo {author} {\bibfnamefont
  {A.}~\bibnamefont {Reyes}},\ and\ \bibinfo {author} {\bibfnamefont {K.~D.}\
  \bibnamefont {Harris}},\ }\href@noop {} {\bibfield  {journal} {\bibinfo
  {journal} {Science}\ }\textbf {\bibinfo {volume} {327}},\ \bibinfo {pages}
  {587} (\bibinfo {year} {2010})}\BibitemShut {NoStop}%
\bibitem [{\citenamefont {Sippy}\ and\ \citenamefont {Yuste}(2013)}]{Yuste}%
  \BibitemOpen
  \bibfield  {author} {\bibinfo {author} {\bibfnamefont {T.}~\bibnamefont
  {Sippy}}\ and\ \bibinfo {author} {\bibfnamefont {R.}~\bibnamefont {Yuste}},\
  }\href {https://doi.org/10.1523/JNEUROSCI.4579-12.2013} {\bibfield  {journal}
  {\bibinfo  {journal} {Journal of Neuroscience}\ }\textbf {\bibinfo {volume}
  {33}},\ \bibinfo {pages} {9813} (\bibinfo {year} {2013})}\BibitemShut
  {NoStop}%
\bibitem [{\citenamefont {Dahmen}\ \emph {et~al.}(2019)\citenamefont {Dahmen},
  \citenamefont {G{r\"u}n}, \citenamefont {Diesmann},\ and\ \citenamefont
  {Helias}}]{Helias}%
  \BibitemOpen
  \bibfield  {author} {\bibinfo {author} {\bibfnamefont {D.}~\bibnamefont
  {Dahmen}}, \bibinfo {author} {\bibfnamefont {S.}~\bibnamefont {G{r\"u}n}},
  \bibinfo {author} {\bibfnamefont {M.}~\bibnamefont {Diesmann}},\ and\
  \bibinfo {author} {\bibfnamefont {M.}~\bibnamefont {Helias}},\ }\href
  {https://doi.org/10.1073/pnas.1818972116} {\bibfield  {journal} {\bibinfo
  {journal} {Proceedings of the National Academy of Sciences}\ }\textbf
  {\bibinfo {volume} {116}},\ \bibinfo {pages} {13051} (\bibinfo {year}
  {2019})}\BibitemShut {NoStop}%
\bibitem [{\citenamefont {Benayoun}\ \emph {et~al.}(2010)\citenamefont
  {Benayoun}, \citenamefont {Cowan}, \citenamefont {van Drongelen},\ and\
  \citenamefont {Wallace}}]{Benayoun}%
  \BibitemOpen
  \bibfield  {author} {\bibinfo {author} {\bibfnamefont {M.}~\bibnamefont
  {Benayoun}}, \bibinfo {author} {\bibfnamefont {J.~D.}\ \bibnamefont {Cowan}},
  \bibinfo {author} {\bibfnamefont {W.}~\bibnamefont {van Drongelen}},\ and\
  \bibinfo {author} {\bibfnamefont {E.}~\bibnamefont {Wallace}},\ }\href@noop
  {} {\bibfield  {journal} {\bibinfo  {journal} {PLoS computational biology}\
  }\textbf {\bibinfo {volume} {6}},\ \bibinfo {pages} {e1000846} (\bibinfo
  {year} {2010})}\BibitemShut {NoStop}%
\bibitem [{\citenamefont {Wilting}\ \emph {et~al.}(2018)\citenamefont
  {Wilting}, \citenamefont {Dehning}, \citenamefont {Pinheiro~Neto},
  \citenamefont {Rudelt}, \citenamefont {Wibral}, \citenamefont {Zierenberg},\
  and\ \citenamefont {Priesemann}}]{Zierenberg18}%
  \BibitemOpen
  \bibfield  {author} {\bibinfo {author} {\bibfnamefont {J.}~\bibnamefont
  {Wilting}}, \bibinfo {author} {\bibfnamefont {J.}~\bibnamefont {Dehning}},
  \bibinfo {author} {\bibfnamefont {J.}~\bibnamefont {Pinheiro~Neto}}, \bibinfo
  {author} {\bibfnamefont {L.}~\bibnamefont {Rudelt}}, \bibinfo {author}
  {\bibfnamefont {M.}~\bibnamefont {Wibral}}, \bibinfo {author} {\bibfnamefont
  {J.}~\bibnamefont {Zierenberg}},\ and\ \bibinfo {author} {\bibfnamefont
  {V.}~\bibnamefont {Priesemann}},\ }\href@noop {} {\bibfield  {journal}
  {\bibinfo  {journal} {Frontiers in Systems Neuroscience}\ }\textbf {\bibinfo
  {volume} {12}},\ \bibinfo {pages} {55} (\bibinfo {year} {2018})}\BibitemShut
  {NoStop}%
\bibitem [{\citenamefont {Liang}\ \emph {et~al.}(2020)\citenamefont {Liang},
  \citenamefont {Zhou},\ and\ \citenamefont {Zhou}}]{Zhou-2020}%
  \BibitemOpen
  \bibfield  {author} {\bibinfo {author} {\bibfnamefont {J.}~\bibnamefont
  {Liang}}, \bibinfo {author} {\bibfnamefont {T.}~\bibnamefont {Zhou}},\ and\
  \bibinfo {author} {\bibfnamefont {C.}~\bibnamefont {Zhou}},\ }\href@noop {}
  {\bibfield  {journal} {\bibinfo  {journal} {Frontiers in systems
  neuroscience}\ }\textbf {\bibinfo {volume} {14}},\ \bibinfo {pages} {87}
  (\bibinfo {year} {2020})}\BibitemShut {NoStop}%
\bibitem [{\citenamefont {Buend{\'\i}a}\ \emph {et~al.}(2019)\citenamefont
  {Buend{\'\i}a}, \citenamefont {Villegas}, \citenamefont {di~Santo},
  \citenamefont {Vezzani}, \citenamefont {Burioni},\ and\ \citenamefont
  {Mu{\~n}oz}}]{Buendia-Jensen}%
  \BibitemOpen
  \bibfield  {author} {\bibinfo {author} {\bibfnamefont {V.}~\bibnamefont
  {Buend{\'\i}a}}, \bibinfo {author} {\bibfnamefont {P.}~\bibnamefont
  {Villegas}}, \bibinfo {author} {\bibfnamefont {S.}~\bibnamefont {di~Santo}},
  \bibinfo {author} {\bibfnamefont {A.}~\bibnamefont {Vezzani}}, \bibinfo
  {author} {\bibfnamefont {R.}~\bibnamefont {Burioni}},\ and\ \bibinfo {author}
  {\bibfnamefont {M.~A.}\ \bibnamefont {Mu{\~n}oz}},\ }\href@noop {} {\bibfield
   {journal} {\bibinfo  {journal} {Scientific Reports}\ }\textbf {\bibinfo
  {volume} {9}} (\bibinfo {year} {2019})}\BibitemShut {NoStop}%
\bibitem [{\citenamefont {Girardi-Schappo}\ \emph {et~al.}(2021)\citenamefont
  {Girardi-Schappo}, \citenamefont {Galera}, \citenamefont {Carvalho},
  \citenamefont {Brochini}, \citenamefont {Kamiji}, \citenamefont {Roque},\
  and\ \citenamefont {Kinouchi}}]{Girardi}%
  \BibitemOpen
  \bibfield  {author} {\bibinfo {author} {\bibfnamefont {M.}~\bibnamefont
  {Girardi-Schappo}}, \bibinfo {author} {\bibfnamefont {E.~F.}\ \bibnamefont
  {Galera}}, \bibinfo {author} {\bibfnamefont {T.~T.}\ \bibnamefont
  {Carvalho}}, \bibinfo {author} {\bibfnamefont {L.}~\bibnamefont {Brochini}},
  \bibinfo {author} {\bibfnamefont {N.~L.}\ \bibnamefont {Kamiji}}, \bibinfo
  {author} {\bibfnamefont {A.~C.}\ \bibnamefont {Roque}},\ and\ \bibinfo
  {author} {\bibfnamefont {O.}~\bibnamefont {Kinouchi}},\ }\href@noop {}
  {\bibfield  {journal} {\bibinfo  {journal} {Journal of Physics: Complexity}\
  }\textbf {\bibinfo {volume} {2}},\ \bibinfo {pages} {045001} (\bibinfo {year}
  {2021})}\BibitemShut {NoStop}%
\bibitem [{\citenamefont {Li}\ and\ \citenamefont {Shew}(2020)}]{Li-Shew}%
  \BibitemOpen
  \bibfield  {author} {\bibinfo {author} {\bibfnamefont {J.}~\bibnamefont
  {Li}}\ and\ \bibinfo {author} {\bibfnamefont {W.~L.}\ \bibnamefont {Shew}},\
  }\href@noop {} {\bibfield  {journal} {\bibinfo  {journal} {PLoS Computational
  Biology}\ }\textbf {\bibinfo {volume} {16}},\ \bibinfo {pages} {e1008268}
  (\bibinfo {year} {2020})}\BibitemShut {NoStop}%
\bibitem [{\citenamefont {Marro}\ and\ \citenamefont {Dickman}(1999)}]{Marro}%
  \BibitemOpen
  \bibfield  {author} {\bibinfo {author} {\bibfnamefont {J.}~\bibnamefont
  {Marro}}\ and\ \bibinfo {author} {\bibfnamefont {R.}~\bibnamefont
  {Dickman}},\ }\href@noop {} {\emph {\bibinfo {title} {Nonequilibrium Phase
  Transition in Lattice Models}}}\ (\bibinfo  {publisher} {Cambridge University
  Press},\ \bibinfo {year} {1999})\BibitemShut {NoStop}%
\bibitem [{\citenamefont {Henkel}\ \emph {et~al.}(2008)\citenamefont {Henkel},
  \citenamefont {Hinrichsen},\ and\ \citenamefont {L{\"u}beck}}]{Henkel}%
  \BibitemOpen
  \bibfield  {author} {\bibinfo {author} {\bibfnamefont {M.}~\bibnamefont
  {Henkel}}, \bibinfo {author} {\bibfnamefont {H.}~\bibnamefont {Hinrichsen}},\
  and\ \bibinfo {author} {\bibfnamefont {S.}~\bibnamefont {L{\"u}beck}},\
  }\href@noop {} {\emph {\bibinfo {title} {Non-equilibrium Phase Transitions:
  Absorbing phase transitions}}},\ Theor. and Math. Phys.\ (\bibinfo
  {publisher} {Springer London},\ \bibinfo {address} {Berlin},\ \bibinfo {year}
  {2008})\BibitemShut {NoStop}%
\bibitem [{\citenamefont {{\'O}dor}(2008)}]{Odor}%
  \BibitemOpen
  \bibfield  {author} {\bibinfo {author} {\bibfnamefont {G.}~\bibnamefont
  {{\'O}dor}},\ }\href@noop {} {\emph {\bibinfo {title} {Universality in
  Nonequilibrium Lattice Systems: Theoretical Foundations}}}\ (\bibinfo
  {publisher} {World Scientific},\ \bibinfo {address} {Singapore},\ \bibinfo
  {year} {2008})\BibitemShut {NoStop}%
\bibitem [{\citenamefont {Gillespie}(2007)}]{Gillespie}%
  \BibitemOpen
  \bibfield  {author} {\bibinfo {author} {\bibfnamefont {D.~T.}\ \bibnamefont
  {Gillespie}},\ }\href@noop {} {\bibfield  {journal} {\bibinfo  {journal}
  {Annu. Rev. Phys. Chem.}\ }\textbf {\bibinfo {volume} {58}},\ \bibinfo
  {pages} {35} (\bibinfo {year} {2007})}\BibitemShut {NoStop}%
\bibitem [{\citenamefont {Pastor-Satorras}\ \emph {et~al.}(2015)\citenamefont
  {Pastor-Satorras}, \citenamefont {Castellano}, \citenamefont {Van~Mieghem},\
  and\ \citenamefont {Vespignani}}]{Romu}%
  \BibitemOpen
  \bibfield  {author} {\bibinfo {author} {\bibfnamefont {R.}~\bibnamefont
  {Pastor-Satorras}}, \bibinfo {author} {\bibfnamefont {C.}~\bibnamefont
  {Castellano}}, \bibinfo {author} {\bibfnamefont {P.}~\bibnamefont
  {Van~Mieghem}},\ and\ \bibinfo {author} {\bibfnamefont {A.}~\bibnamefont
  {Vespignani}},\ }\href@noop {} {\bibfield  {journal} {\bibinfo  {journal}
  {Reviews of modern physics}\ }\textbf {\bibinfo {volume} {87}},\ \bibinfo
  {pages} {925} (\bibinfo {year} {2015})}\BibitemShut {NoStop}%
\bibitem [{\citenamefont {Cowan}\ \emph {et~al.}(2016)\citenamefont {Cowan},
  \citenamefont {Neuman},\ and\ \citenamefont {van Drongelen}}]{WC-review}%
  \BibitemOpen
  \bibfield  {author} {\bibinfo {author} {\bibfnamefont {J.~D.}\ \bibnamefont
  {Cowan}}, \bibinfo {author} {\bibfnamefont {J.}~\bibnamefont {Neuman}},\ and\
  \bibinfo {author} {\bibfnamefont {W.}~\bibnamefont {van Drongelen}},\
  }\href@noop {} {\bibfield  {journal} {\bibinfo  {journal} {The Journal of
  Mathematical Neuroscience}\ }\textbf {\bibinfo {volume} {6}},\ \bibinfo
  {pages} {1} (\bibinfo {year} {2016})}\BibitemShut {NoStop}%
\bibitem [{\citenamefont {Negahbani}\ \emph {et~al.}(2015)\citenamefont
  {Negahbani}, \citenamefont {Steyn-Ross}, \citenamefont {Steyn-Ross},
  \citenamefont {Wilson},\ and\ \citenamefont {Sleigh}}]{Negah}%
  \BibitemOpen
  \bibfield  {author} {\bibinfo {author} {\bibfnamefont {E.}~\bibnamefont
  {Negahbani}}, \bibinfo {author} {\bibfnamefont {D.~A.}\ \bibnamefont
  {Steyn-Ross}}, \bibinfo {author} {\bibfnamefont {M.~L.}\ \bibnamefont
  {Steyn-Ross}}, \bibinfo {author} {\bibfnamefont {M.~T.}\ \bibnamefont
  {Wilson}},\ and\ \bibinfo {author} {\bibfnamefont {J.~W.}\ \bibnamefont
  {Sleigh}},\ }\href@noop {} {\bibfield  {journal} {\bibinfo  {journal} {J. of
  Math. Neurosci}\ }\textbf {\bibinfo {volume} {5}},\ \bibinfo {pages} {1}
  (\bibinfo {year} {2015})}\BibitemShut {NoStop}%
\bibitem [{\citenamefont {di~Santo}\ \emph {et~al.}(2018)\citenamefont
  {di~Santo}, \citenamefont {Villegas}, \citenamefont {Burioni},\ and\
  \citenamefont {Mu{\~n}oz}}]{Serena-NN}%
  \BibitemOpen
  \bibfield  {author} {\bibinfo {author} {\bibfnamefont {S.}~\bibnamefont
  {di~Santo}}, \bibinfo {author} {\bibfnamefont {P.}~\bibnamefont {Villegas}},
  \bibinfo {author} {\bibfnamefont {R.}~\bibnamefont {Burioni}},\ and\ \bibinfo
  {author} {\bibfnamefont {M.~A.}\ \bibnamefont {Mu{\~n}oz}},\ }\href@noop {}
  {\bibfield  {journal} {\bibinfo  {journal} {Journal of Statistical Mechanics:
  Theory and Experiment}\ }\textbf {\bibinfo {volume} {2018}},\ \bibinfo
  {pages} {073402} (\bibinfo {year} {2018})}\BibitemShut {NoStop}%
\bibitem [{\citenamefont {De~Candia}\ \emph {et~al.}(2021)\citenamefont
  {De~Candia}, \citenamefont {Sarracino}, \citenamefont {Apicella},\ and\
  \citenamefont {de~Arcangelis}}]{Lucilla}%
  \BibitemOpen
  \bibfield  {author} {\bibinfo {author} {\bibfnamefont {A.}~\bibnamefont
  {De~Candia}}, \bibinfo {author} {\bibfnamefont {A.}~\bibnamefont
  {Sarracino}}, \bibinfo {author} {\bibfnamefont {I.}~\bibnamefont
  {Apicella}},\ and\ \bibinfo {author} {\bibfnamefont {L.}~\bibnamefont
  {de~Arcangelis}},\ }\href
  {https://doi.org/doi.org/10.1371/journal.pcbi.1008884} {\bibfield  {journal}
  {\bibinfo  {journal} {PLoS Comput Biol}\ ,\ \bibinfo {pages} {e1008884}}
  (\bibinfo {year} {2021})}\BibitemShut {NoStop}%
\bibitem [{\citenamefont {Sarracino}\ \emph {et~al.}(2020)\citenamefont
  {Sarracino}, \citenamefont {Arviv}, \citenamefont {Shriki},\ and\
  \citenamefont {De~Arcangelis}}]{Lucilla2}%
  \BibitemOpen
  \bibfield  {author} {\bibinfo {author} {\bibfnamefont {A.}~\bibnamefont
  {Sarracino}}, \bibinfo {author} {\bibfnamefont {O.}~\bibnamefont {Arviv}},
  \bibinfo {author} {\bibfnamefont {O.}~\bibnamefont {Shriki}},\ and\ \bibinfo
  {author} {\bibfnamefont {L.}~\bibnamefont {De~Arcangelis}},\ }\href@noop {}
  {\bibfield  {journal} {\bibinfo  {journal} {Physical Review Research}\
  }\textbf {\bibinfo {volume} {2}},\ \bibinfo {pages} {033355} (\bibinfo {year}
  {2020})}\BibitemShut {NoStop}%
\bibitem [{\citenamefont {Kunze}(2000)}]{Kunze}%
  \BibitemOpen
  \bibfield  {author} {\bibinfo {author} {\bibfnamefont {M.}~\bibnamefont
  {Kunze}},\ }\href@noop {} {\emph {\bibinfo {title} {Non-smooth dynamical
  systems}}},\ Vol.\ \bibinfo {volume} {1744}\ (\bibinfo  {publisher} {Springer
  Science \& Business Media},\ \bibinfo {year} {2000})\BibitemShut {NoStop}%
\bibitem [{\citenamefont {Izhikevich}(2007)}]{Izhi}%
  \BibitemOpen
  \bibfield  {author} {\bibinfo {author} {\bibfnamefont {E.~M.}\ \bibnamefont
  {Izhikevich}},\ }\href@noop {} {\emph {\bibinfo {title} {Dynamical systems in
  neuroscience}}}\ (\bibinfo  {publisher} {MIT press},\ \bibinfo {year}
  {2007})\BibitemShut {NoStop}%
\bibitem [{\citenamefont {Murphy}\ and\ \citenamefont
  {Miller}(2009)}]{MurphyMiller}%
  \BibitemOpen
  \bibfield  {author} {\bibinfo {author} {\bibfnamefont {B.~K.}\ \bibnamefont
  {Murphy}}\ and\ \bibinfo {author} {\bibfnamefont {K.~D.}\ \bibnamefont
  {Miller}},\ }\href@noop {} {\bibfield  {journal} {\bibinfo  {journal}
  {Neuron}\ }\textbf {\bibinfo {volume} {61}},\ \bibinfo {pages} {635}
  (\bibinfo {year} {2009})}\BibitemShut {NoStop}%
\bibitem [{\citenamefont {Hennequin}\ \emph {et~al.}(2012)\citenamefont
  {Hennequin}, \citenamefont {Vogels},\ and\ \citenamefont
  {Gerstner}}]{Gertsner}%
  \BibitemOpen
  \bibfield  {author} {\bibinfo {author} {\bibfnamefont {G.}~\bibnamefont
  {Hennequin}}, \bibinfo {author} {\bibfnamefont {T.~P.}\ \bibnamefont
  {Vogels}},\ and\ \bibinfo {author} {\bibfnamefont {W.}~\bibnamefont
  {Gerstner}},\ }\href@noop {} {\bibfield  {journal} {\bibinfo  {journal}
  {Physical Review E}\ }\textbf {\bibinfo {volume} {86}},\ \bibinfo {pages}
  {011909} (\bibinfo {year} {2012})}\BibitemShut {NoStop}%
\bibitem [{\citenamefont {Fruchart}\ \emph {et~al.}(2021)\citenamefont
  {Fruchart}, \citenamefont {Hanai}, \citenamefont {Littlewood},\ and\
  \citenamefont {Vitelli}}]{Non-reciprocal}%
  \BibitemOpen
  \bibfield  {author} {\bibinfo {author} {\bibfnamefont {M.}~\bibnamefont
  {Fruchart}}, \bibinfo {author} {\bibfnamefont {R.}~\bibnamefont {Hanai}},
  \bibinfo {author} {\bibfnamefont {P.~B.}\ \bibnamefont {Littlewood}},\ and\
  \bibinfo {author} {\bibfnamefont {V.}~\bibnamefont {Vitelli}},\ }\href@noop
  {} {\bibfield  {journal} {\bibinfo  {journal} {Nature}\ }\textbf {\bibinfo
  {volume} {592}},\ \bibinfo {pages} {363} (\bibinfo {year}
  {2021})}\BibitemShut {NoStop}%
\bibitem [{\citenamefont {Ponce-Alvarez}\ \emph
  {et~al.}(2018{\natexlab{b}})\citenamefont {Ponce-Alvarez}, \citenamefont
  {Jouary}, \citenamefont {Privat}, \citenamefont {Deco},\ and\ \citenamefont
  {Sumbre}}]{zebra}%
  \BibitemOpen
  \bibfield  {author} {\bibinfo {author} {\bibfnamefont {A.}~\bibnamefont
  {Ponce-Alvarez}}, \bibinfo {author} {\bibfnamefont {A.}~\bibnamefont
  {Jouary}}, \bibinfo {author} {\bibfnamefont {M.}~\bibnamefont {Privat}},
  \bibinfo {author} {\bibfnamefont {G.}~\bibnamefont {Deco}},\ and\ \bibinfo
  {author} {\bibfnamefont {G.}~\bibnamefont {Sumbre}},\ }\href@noop {}
  {\bibfield  {journal} {\bibinfo  {journal} {Neuron}\ }\textbf {\bibinfo
  {volume} {100}},\ \bibinfo {pages} {1446} (\bibinfo {year}
  {2018}{\natexlab{b}})}\BibitemShut {NoStop}%
\bibitem [{\citenamefont {Piuvezam}\ and\ \citenamefont
  {others.}(2022)}]{Helena}%
  \BibitemOpen
  \bibfield  {author} {\bibinfo {author} {\bibfnamefont {H.}~\bibnamefont
  {Piuvezam}}\ and\ \bibinfo {author} {\bibnamefont {others.}},\ }\href@noop {}
  {\bibfield  {journal} {\bibinfo  {journal} {preprint}\ } (\bibinfo {year}
  {2022})}\BibitemShut {NoStop}%
\bibitem [{\citenamefont {Li}\ \emph {et~al.}(2009)\citenamefont {Li},
  \citenamefont {Poo},\ and\ \citenamefont {Dan}}]{Yang-Dan}%
  \BibitemOpen
  \bibfield  {author} {\bibinfo {author} {\bibfnamefont {C.-y.~T.}\
  \bibnamefont {Li}}, \bibinfo {author} {\bibfnamefont {M.-m.}\ \bibnamefont
  {Poo}},\ and\ \bibinfo {author} {\bibfnamefont {Y.}~\bibnamefont {Dan}},\
  }\href@noop {} {\bibfield  {journal} {\bibinfo  {journal} {Science}\ }\textbf
  {\bibinfo {volume} {324}},\ \bibinfo {pages} {643} (\bibinfo {year}
  {2009})}\BibitemShut {NoStop}%
\bibitem [{\citenamefont {Hansel}\ and\ \citenamefont
  {Mato}(2001)}]{Hansel-Mato}%
  \BibitemOpen
  \bibfield  {author} {\bibinfo {author} {\bibfnamefont {D.}~\bibnamefont
  {Hansel}}\ and\ \bibinfo {author} {\bibfnamefont {G.}~\bibnamefont {Mato}},\
  }\href@noop {} {\bibfield  {journal} {\bibinfo  {journal} {Physical Review
  Letters}\ }\textbf {\bibinfo {volume} {86}},\ \bibinfo {pages} {4175}
  (\bibinfo {year} {2001})}\BibitemShut {NoStop}%
\bibitem [{\citenamefont {Kriener}\ \emph {et~al.}(2014)\citenamefont
  {Kriener}, \citenamefont {Enger}, \citenamefont {Tetzlaff}, \citenamefont
  {Plesser}, \citenamefont {Gewaltig},\ and\ \citenamefont
  {Einevoll}}]{Einevoll}%
  \BibitemOpen
  \bibfield  {author} {\bibinfo {author} {\bibfnamefont {B.}~\bibnamefont
  {Kriener}}, \bibinfo {author} {\bibfnamefont {H.}~\bibnamefont {Enger}},
  \bibinfo {author} {\bibfnamefont {T.}~\bibnamefont {Tetzlaff}}, \bibinfo
  {author} {\bibfnamefont {H.~E.}\ \bibnamefont {Plesser}}, \bibinfo {author}
  {\bibfnamefont {M.-O.}\ \bibnamefont {Gewaltig}},\ and\ \bibinfo {author}
  {\bibfnamefont {G.~T.}\ \bibnamefont {Einevoll}},\ }\href@noop {} {\bibfield
  {journal} {\bibinfo  {journal} {Frontiers in computational neuroscience}\
  }\textbf {\bibinfo {volume} {8}},\ \bibinfo {pages} {136} (\bibinfo {year}
  {2014})}\BibitemShut {NoStop}%
\bibitem [{\citenamefont {Asllani}\ \emph {et~al.}(2018)\citenamefont
  {Asllani}, \citenamefont {Lambiotte},\ and\ \citenamefont
  {Carletti}}]{Lambiotte}%
  \BibitemOpen
  \bibfield  {author} {\bibinfo {author} {\bibfnamefont {M.}~\bibnamefont
  {Asllani}}, \bibinfo {author} {\bibfnamefont {R.}~\bibnamefont {Lambiotte}},\
  and\ \bibinfo {author} {\bibfnamefont {T.}~\bibnamefont {Carletti}},\
  }\href@noop {} {\bibfield  {journal} {\bibinfo  {journal} {Science advances}\
  }\textbf {\bibinfo {volume} {4}},\ \bibinfo {pages} {eaau9403} (\bibinfo
  {year} {2018})}\BibitemShut {NoStop}%
\bibitem [{\citenamefont {Durstewitz}\ \emph {et~al.}(2000)\citenamefont
  {Durstewitz}, \citenamefont {Seamans},\ and\ \citenamefont {Sejnowski}}]{WM}%
  \BibitemOpen
  \bibfield  {author} {\bibinfo {author} {\bibfnamefont {D.}~\bibnamefont
  {Durstewitz}}, \bibinfo {author} {\bibfnamefont {J.~K.}\ \bibnamefont
  {Seamans}},\ and\ \bibinfo {author} {\bibfnamefont {T.~J.}\ \bibnamefont
  {Sejnowski}},\ }\href@noop {} {\bibfield  {journal} {\bibinfo  {journal}
  {Nature neuroscience}\ }\textbf {\bibinfo {volume} {3}},\ \bibinfo {pages}
  {1184} (\bibinfo {year} {2000})}\BibitemShut {NoStop}%
\bibitem [{\citenamefont {Wallace}\ \emph {et~al.}(2011)\citenamefont
  {Wallace}, \citenamefont {Benayoun}, \citenamefont {van Drongelen},\ and\
  \citenamefont {Cowan}}]{Wallace}%
  \BibitemOpen
  \bibfield  {author} {\bibinfo {author} {\bibfnamefont {E.}~\bibnamefont
  {Wallace}}, \bibinfo {author} {\bibfnamefont {M.}~\bibnamefont {Benayoun}},
  \bibinfo {author} {\bibfnamefont {W.}~\bibnamefont {van Drongelen}},\ and\
  \bibinfo {author} {\bibfnamefont {J.}~\bibnamefont {Cowan}},\ }\href@noop {}
  {\bibfield  {journal} {\bibinfo  {journal} {PLoS One}\ }\textbf {\bibinfo
  {volume} {6}},\ \bibinfo {pages} {e14804} (\bibinfo {year}
  {2011})}\BibitemShut {NoStop}%
\bibitem [{\citenamefont {Hidalgo}\ \emph {et~al.}(2012)\citenamefont
  {Hidalgo}, \citenamefont {Seoane}, \citenamefont {Cort\'es},\ and\
  \citenamefont {Mu{\~n}oz}}]{Hidalgo}%
  \BibitemOpen
  \bibfield  {author} {\bibinfo {author} {\bibfnamefont {J.}~\bibnamefont
  {Hidalgo}}, \bibinfo {author} {\bibfnamefont {L.}~\bibnamefont {Seoane}},
  \bibinfo {author} {\bibfnamefont {J.}~\bibnamefont {Cort\'es}},\ and\
  \bibinfo {author} {\bibfnamefont {M.}~\bibnamefont {Mu{\~n}oz}},\ }\href@noop
  {} {\bibfield  {journal} {\bibinfo  {journal} {PLoS One}\ }\textbf {\bibinfo
  {volume} {7(8)}},\ \bibinfo {pages} {e40710} (\bibinfo {year}
  {2012})}\BibitemShut {NoStop}%
\bibitem [{\citenamefont {Kim}\ and\ \citenamefont
  {Sejnowski}(2021)}]{Sejnowski}%
  \BibitemOpen
  \bibfield  {author} {\bibinfo {author} {\bibfnamefont {R.}~\bibnamefont
  {Kim}}\ and\ \bibinfo {author} {\bibfnamefont {T.~J.}\ \bibnamefont
  {Sejnowski}},\ }\href@noop {} {\bibfield  {journal} {\bibinfo  {journal}
  {Nature Neuroscience}\ }\textbf {\bibinfo {volume} {24}},\ \bibinfo {pages}
  {129} (\bibinfo {year} {2021})}\BibitemShut {NoStop}%
\bibitem [{\citenamefont {Gleeson}\ and\ \citenamefont
  {Durrett}(2017)}]{Gleeson}%
  \BibitemOpen
  \bibfield  {author} {\bibinfo {author} {\bibfnamefont {J.~P.}\ \bibnamefont
  {Gleeson}}\ and\ \bibinfo {author} {\bibfnamefont {R.}~\bibnamefont
  {Durrett}},\ }\href@noop {} {\bibfield  {journal} {\bibinfo  {journal}
  {Nature Communications}\ }\textbf {\bibinfo {volume} {8}},\ \bibinfo {pages}
  {1227} (\bibinfo {year} {2017})}\BibitemShut {NoStop}%
\end{thebibliography}%


\begin{thebibliography}{12}%
\makeatletter
\providecommand \@ifxundefined [1]{%
 \@ifx{#1\undefined}
}%
\providecommand \@ifnum [1]{%
 \ifnum #1\expandafter \@firstoftwo
 \else \expandafter \@secondoftwo
 \fi
}%
\providecommand \@ifx [1]{%
 \ifx #1\expandafter \@firstoftwo
 \else \expandafter \@secondoftwo
 \fi
}%
\providecommand \natexlab [1]{#1}%
\providecommand \enquote  [1]{``#1''}%
\providecommand \bibnamefont  [1]{#1}%
\providecommand \bibfnamefont [1]{#1}%
\providecommand \citenamefont [1]{#1}%
\providecommand \href@noop [0]{\@secondoftwo}%
\providecommand \href [0]{\begingroup \@sanitize@url \@href}%
\providecommand \@href[1]{\@@startlink{#1}\@@href}%
\providecommand \@@href[1]{\endgroup#1\@@endlink}%
\providecommand \@sanitize@url [0]{\catcode `\\12\catcode `\$12\catcode
  `\&12\catcode `\#12\catcode `\^12\catcode `\_12\catcode `\%12\relax}%
\providecommand \@@startlink[1]{}%
\providecommand \@@endlink[0]{}%
\providecommand \url  [0]{\begingroup\@sanitize@url \@url }%
\providecommand \@url [1]{\endgroup\@href {#1}{\urlprefix }}%
\providecommand \urlprefix  [0]{URL }%
\providecommand \Eprint [0]{\href }%
\providecommand \doibase [0]{https://doi.org/}%
\providecommand \selectlanguage [0]{\@gobble}%
\providecommand \bibinfo  [0]{\@secondoftwo}%
\providecommand \bibfield  [0]{\@secondoftwo}%
\providecommand \translation [1]{[#1]}%
\providecommand \BibitemOpen [0]{}%
\providecommand \bibitemStop [0]{}%
\providecommand \bibitemNoStop [0]{.\EOS\space}%
\providecommand \EOS [0]{\spacefactor3000\relax}%
\providecommand \BibitemShut  [1]{\csname bibitem#1\endcsname}%
\let\auto@bib@innerbib\@empty
%</preamble>
\bibitem [{\citenamefont {Hinrichsen}(2000)}]{Hinrichsen}%
  \BibitemOpen
  \bibfield  {author} {\bibinfo {author} {\bibfnamefont {H.}~\bibnamefont
  {Hinrichsen}},\ }\bibfield  {title} {\bibinfo {title} {Non-equilibrium
  critical phenomena and phase transitions into absorbing states},\ }\href@noop
  {} {\bibfield  {journal} {\bibinfo  {journal} {Adv. in Phys.}\ }\textbf
  {\bibinfo {volume} {49}},\ \bibinfo {pages} {815} (\bibinfo {year}
  {2000})}\BibitemShut {NoStop}%
\bibitem [{\citenamefont {Gillespie}(2007)}]{Gillespie}%
  \BibitemOpen
  \bibfield  {author} {\bibinfo {author} {\bibfnamefont {D.~T.}\ \bibnamefont
  {Gillespie}},\ }\bibfield  {title} {\bibinfo {title} {Stochastic simulation
  of chemical kinetics},\ }\href@noop {} {\bibfield  {journal} {\bibinfo
  {journal} {Annu. Rev. Phys. Chem.}\ }\textbf {\bibinfo {volume} {58}},\
  \bibinfo {pages} {35} (\bibinfo {year} {2007})}\BibitemShut {NoStop}%
\bibitem [{\citenamefont {Gardiner}(2009)}]{Gardiner}%
  \BibitemOpen
  \bibfield  {author} {\bibinfo {author} {\bibfnamefont {C.}~\bibnamefont
  {Gardiner}},\ }\href {http://books.google.it/books?id=otg3PQAACAAJ} {\emph
  {\bibinfo {title} {Stochastic Methods: A Handbook for the Natural and Social
  Sciences}}},\ Springer Series in Synergetics\ (\bibinfo  {publisher}
  {Springer},\ \bibinfo {year} {2009})\BibitemShut {NoStop}%
\bibitem [{\citenamefont {Strogatz}(2018)}]{Strogatz}%
  \BibitemOpen
  \bibfield  {author} {\bibinfo {author} {\bibfnamefont {S.~H.}\ \bibnamefont
  {Strogatz}},\ }\href@noop {} {\emph {\bibinfo {title} {Nonlinear dynamics and
  chaos with student solutions manual: With applications to physics, biology,
  chemistry, and engineering}}}\ (\bibinfo  {publisher} {CRC press},\ \bibinfo
  {year} {2018})\BibitemShut {NoStop}%
\bibitem [{\citenamefont {Buend{\'\i}a}\ \emph {et~al.}(2019)\citenamefont
  {Buend{\'\i}a}, \citenamefont {Villegas}, \citenamefont {di~Santo},
  \citenamefont {Vezzani}, \citenamefont {Burioni},\ and\ \citenamefont
  {Mu{\~n}oz}}]{Buendia-Jensen}%
  \BibitemOpen
  \bibfield  {author} {\bibinfo {author} {\bibfnamefont {V.}~\bibnamefont
  {Buend{\'\i}a}}, \bibinfo {author} {\bibfnamefont {P.}~\bibnamefont
  {Villegas}}, \bibinfo {author} {\bibfnamefont {S.}~\bibnamefont {di~Santo}},
  \bibinfo {author} {\bibfnamefont {A.}~\bibnamefont {Vezzani}}, \bibinfo
  {author} {\bibfnamefont {R.}~\bibnamefont {Burioni}},\ and\ \bibinfo {author}
  {\bibfnamefont {M.~A.}\ \bibnamefont {Mu{\~n}oz}},\ }\bibfield  {title}
  {\bibinfo {title} {Jensen’s force and the statistical mechanics of cortical
  asynchronous states},\ }\href@noop {} {\bibfield  {journal} {\bibinfo
  {journal} {Scientific Reports}\ }\textbf {\bibinfo {volume} {9}} (\bibinfo
  {year} {2019})}\BibitemShut {NoStop}%
\bibitem [{\citenamefont {Steriade}\ \emph {et~al.}(1993)\citenamefont
  {Steriade}, \citenamefont {Nunez},\ and\ \citenamefont {Amzica}}]{Steriade}%
  \BibitemOpen
  \bibfield  {author} {\bibinfo {author} {\bibfnamefont {M.}~\bibnamefont
  {Steriade}}, \bibinfo {author} {\bibfnamefont {A.}~\bibnamefont {Nunez}},\
  and\ \bibinfo {author} {\bibfnamefont {F.}~\bibnamefont {Amzica}},\
  }\bibfield  {title} {\bibinfo {title} {A novel slow oscillation of
  neocortical neurons in vivo: depolarizing and hyperpolarizing components},\
  }\href@noop {} {\bibfield  {journal} {\bibinfo  {journal} {Journal of
  Neurosci.}\ }\textbf {\bibinfo {volume} {13}},\ \bibinfo {pages} {3252}
  (\bibinfo {year} {1993})}\BibitemShut {NoStop}%
\bibitem [{\citenamefont {Destexhe}(2009)}]{Destexhe}%
  \BibitemOpen
  \bibfield  {author} {\bibinfo {author} {\bibfnamefont {A.}~\bibnamefont
  {Destexhe}},\ }\bibfield  {title} {\bibinfo {title} {Self-sustained
  asynchronous irregular states and up–down states in thalamic, cortical and
  thalamocortical networks of nonlinear integrate-and-fire neurons},\
  }\href@noop {} {\bibfield  {journal} {\bibinfo  {journal} {Journal of
  computational neuroscience}\ }\textbf {\bibinfo {volume} {27}},\ \bibinfo
  {pages} {493} (\bibinfo {year} {2009})}\BibitemShut {NoStop}%
\bibitem [{\citenamefont {Marro}\ and\ \citenamefont {Dickman}(1999)}]{Marro}%
  \BibitemOpen
  \bibfield  {author} {\bibinfo {author} {\bibfnamefont {J.}~\bibnamefont
  {Marro}}\ and\ \bibinfo {author} {\bibfnamefont {R.}~\bibnamefont
  {Dickman}},\ }\href@noop {} {\emph {\bibinfo {title} {Nonequilibrium Phase
  Transition in Lattice Models}}}\ (\bibinfo  {publisher} {Cambridge University
  Press},\ \bibinfo {year} {1999})\BibitemShut {NoStop}%
\bibitem [{\citenamefont {Henkel}\ \emph {et~al.}(2008)\citenamefont {Henkel},
  \citenamefont {Hinrichsen},\ and\ \citenamefont {L{\"u}beck}}]{Henkel}%
  \BibitemOpen
  \bibfield  {author} {\bibinfo {author} {\bibfnamefont {M.}~\bibnamefont
  {Henkel}}, \bibinfo {author} {\bibfnamefont {H.}~\bibnamefont {Hinrichsen}},\
  and\ \bibinfo {author} {\bibfnamefont {S.}~\bibnamefont {L{\"u}beck}},\
  }\href@noop {} {\emph {\bibinfo {title} {Non-equilibrium Phase Transitions:
  Absorbing phase transitions}}},\ Theor. and Math. Phys.\ (\bibinfo
  {publisher} {Springer London},\ \bibinfo {address} {Berlin},\ \bibinfo {year}
  {2008})\BibitemShut {NoStop}%
\bibitem [{\citenamefont {Mu{\~n}oz}\ \emph {et~al.}(1999)\citenamefont
  {Mu{\~n}oz}, \citenamefont {Dickman}, \citenamefont {Vespignani},\ and\
  \citenamefont {Zapperi}}]{CP-Exponents}%
  \BibitemOpen
  \bibfield  {author} {\bibinfo {author} {\bibfnamefont {M.~A.}\ \bibnamefont
  {Mu{\~n}oz}}, \bibinfo {author} {\bibfnamefont {R.}~\bibnamefont {Dickman}},
  \bibinfo {author} {\bibfnamefont {A.}~\bibnamefont {Vespignani}},\ and\
  \bibinfo {author} {\bibfnamefont {S.}~\bibnamefont {Zapperi}},\ }\bibfield
  {title} {\bibinfo {title} {Avalanche and spreading exponents in systems with
  absorbing states},\ }\href@noop {} {\bibfield  {journal} {\bibinfo  {journal}
  {Phys. Rev. E}\ }\textbf {\bibinfo {volume} {59}},\ \bibinfo {pages} {6175}
  (\bibinfo {year} {1999})}\BibitemShut {NoStop}%
\bibitem [{\citenamefont {Trefethen}\ \emph {et~al.}(2005)\citenamefont
  {Trefethen}, \citenamefont {Embree},\ and\ \citenamefont
  {Embree}}]{PseudoSpectra}%
  \BibitemOpen
  \bibfield  {author} {\bibinfo {author} {\bibfnamefont {L.}~\bibnamefont
  {Trefethen}}, \bibinfo {author} {\bibfnamefont {M.}~\bibnamefont {Embree}},\
  and\ \bibinfo {author} {\bibfnamefont {M.}~\bibnamefont {Embree}},\ }\href
  {https://books.google.es/books?id=7gIbT-Y7-AIC} {\emph {\bibinfo {title}
  {Spectra and Pseudospectra: The Behavior of Nonnormal Matrices and
  Operators}}}\ (\bibinfo  {publisher} {Princeton University Press},\ \bibinfo
  {year} {2005})\BibitemShut {NoStop}%
\bibitem [{\citenamefont {Elsner}\ and\ \citenamefont
  {Paardekooper}(1987)}]{NonNormalityMeasures}%
  \BibitemOpen
  \bibfield  {author} {\bibinfo {author} {\bibfnamefont {L.}~\bibnamefont
  {Elsner}}\ and\ \bibinfo {author} {\bibfnamefont {M.}~\bibnamefont
  {Paardekooper}},\ }\bibfield  {title} {\bibinfo {title} {On measures of
  nonnormality of matrices},\ }\href@noop {} {\bibfield  {journal} {\bibinfo
  {journal} {Linear Algebra and its Applications}\ }\textbf {\bibinfo {volume}
  {92}},\ \bibinfo {pages} {107} (\bibinfo {year} {1987})}\BibitemShut
  {NoStop}%
\end{thebibliography}%

%\end{thebibliography}
\end{document}

\title{The excitatory-inhibitory branching process: a parsimonious
	view of cortical asynchronous states, excitability, and criticality.}

\author{Roberto Corral L\'opez}

\affiliation{Departamento de
	Electromagnetismo y F{\'\i}sica de la Materia e Instituto Carlos I
	de F{\'\i}sica Te{\'o}rica y Computacional. Universidad de Granada.
	E-18071, Granada, Spain}
\author{V{\'\i}ctor Buend{\'\i}a} 
\affiliation{Department of Computer
	Science, University of T{\"u}bingen, and Max Planck Institute for
	Biological Cybernetics, T{\"u}bingen, 72076, Germany}
\author{Miguel A. Mu\~noz} \affiliation{Departamento de
	Electromagnetismo y F{\'\i}sica de la Materia e Instituto Carlos I
	de F{\'\i}sica Te{\'o}rica y Computacional. Universidad de Granada.
	E-18071, Granada, Spain}
	
	\maketitle

	\section{Derivation of the model \label{sec:derivation}}
	
	The EI-CP is a Markov model, thus completely defined by the transition rates between different configurations. At any time $t$ each node in a network $j$ can be either active or inactive --represented by $s_j=1$ and $s_j=0$, respectively-- and the system configuration can be written as  $\vec s = \{ s_0, \ldots, s_N \}$. $N$ is the network size of which $N_e=\alpha N$ are excitatory nodes and $N_i=(1-\alpha)N$  inhibitory. An active node can spontaneously decay to the inactive state at a constant rate $\mu_{e,i}$, where the subindex stands for excitatory and inhibitory types, respectively. The rate at which a node becomes active depends on its type, as well as on the state of the neighbouring nodes. For an excitatory node, the rate is computed by  (i) summing the number of active excitatory neighbours, multiplied by the rate $\lambda$ associated at each interaction (ii) substracting the number of inhibitory neighbours multiplied by $\lambda r_e$, and (iii) filtering the result (divided by the total number of neighbours) through a transfer (sigmoidal) function $f:\mathbb{R}\to\mathbb{R}^+_0$ which ensures that rates are positive, despite of any possible inhibitory effect. The setup is the same for inhibitory individuals, but using $r_i$ instead.  More formally, the transition rates can be generically written as:
	
	\begin{subequations}
	\begin{align}
	\omega^{e/i}_j(1\to 0) &= \mu_{e/i}, \\
	\omega^{e/i}_j(0\to 1) &= f\left(\frac{\lambda}{K_j}\sum_{k\in\Omega_j^ e}s_k-\frac{\lambda r_{e/i}}{K_j}\sum_{k\in\Omega_j^i}s_k\right).
	\end{align}
	\end{subequations}

	where $K_j$ is the connectivity of the node $j$, $f(\Lambda)\equiv \max(\Lambda,0)$ is the usual rectified linear function, $r_{e/i}$ the strength of inhibition over the activation of excitatory/inhibitory units and $\Omega_j^{e,i}$ the set of excitatory/inhibitory neighbours of node $j$. The notation $\omega_j(s\to s')$ can be seen as shorthand for $\omega(\{s_0, \ldots, s_j=s, \ldots, s_N \} \to \{s_0, \ldots, s_j=s', \ldots, s_N \} )$. The evolution of the probability distribution of finding a specific configuration $\vec s$ at time $t$, $P(\vec s, t)$,  can be obtained using the master equation, 
	
	\begin{equation} \label{eqn:FullMasterEq}
		\begin{split}
			\frac{dP(\vec{s}, t)}{dt}=&\sum_{j\in\text{exc}}(2s_j-1)\left\{-\mu_eP(s_1,\ldots,\underset{j}{1},\ldots,s_N;t)\right.\\
			&\left.+f\left(\frac{\lambda}{K_j}\sum_{k\in\Omega_j^e}s_k-\frac{\lambda r_e}{K_j}\sum_{k\in\Omega_j^i}s_k\right)P(s_1, \ldots, \underset{k}{1},\ldots,\underset{j}{0},\ldots,s_N; t)
			\right\}\\
			&+\sum_{j\in\text{inh}}(2s_j-1)\left\{-\mu_iP(s_1,\ldots,\underset{j}{1},\ldots,s_N;t)\right.\\
			&\left.+f\left(\frac{\lambda}{K_j}\sum_{k\in\Omega_j^e}s_k-\frac{\lambda r_i}{K_j}\sum_{k\in\Omega_j^i}s_k\right)P(s_1, \ldots, \underset{k}{1},\ldots,\underset{j}{0},\ldots,s_N; t)
			\right\}, \\
		\end{split}
	\end{equation}
	
	which is formally identical to the Master equation of the ordinary contact process (CP) with the addition of inhibitory nodes and the modification on the activation rates \cite{Hinrichsen}. Numerical simulations are performed integrating the above equation with the  Gillespie algorithm, which is exact \cite{Gillespie}. 
	
	Although the Master equation contains all the information about the system, it is analytically intractable. However, we are interested only in macroscopic variables such as the average density of active units or \emph{activity}. Therefore, it is possible to define the number of active excitatory nodes at a certain time as $n_e(t) = \sum _{j\in \text{exc}} s_j(t)$ (analogously for the inhibitory ones). If we consider that the system is characterized by $\vec n = \{n_e, n_i\}$, then the transition rates are given by
	
	\begin{subequations}
	\begin{align}
	\omega(\{n_e, n_i\}\to\{n_e-1, n_i\})&=\sum_{j\in\text{exc}}\mu_es_j=\mu_e\alpha N, \\
	\omega(\{n_e, n_i\}\to\{n_e, n_i-1\})&=\sum_{j\in\text{inh}}\mu_is_j=\mu_i(1-\alpha)N, \\
	\omega(\{n_e, n_i\}\to\{n_e+1, n_i\})&=\sum_{j\in\text{exc}}(1-s_j)f\left(\frac{\lambda}{K_j}\sum_{k\in\Omega_j^ e}s_k-\frac{\lambda r_e}{K_j}\sum_{k\in\Omega_j^i}s_k\right),\\
	\omega(\{n_e, n_i\}\to\{n_e, n_i+1\})&=\sum_{j\in\text{inh}}(1-s_j)f\left(\frac{\lambda}{K_j}\sum_{k\in\Omega_j^ e}s_k-\frac{\lambda r_i}{K_j}\sum_{k\in\Omega_j^i}s_k\right).
	\end{align}
	\end{subequations}
	
	 As before, from this rates we can construct a Master equation for the new macroscopic variables, which would render the evolution of $P(\vec n, t)$. Now, as customarily done, one can derive the associated Fokker-Planck equation by means of the Kramers-Moyal expansion \cite{Gardiner}. In order to do so, we first need to make the creation rates independent of the microscopic structure by approximating the rate of node activation as the amount of free sites multiplied by the average probability rate of becoming active,
	 
	\begin{subequations}
	\begin{align}
		\omega(\{n_e, n_i\}\to\{n_e+1, n_i\})&\approx(N_e-n_e)\langle f(\Lambda_e)\rangle, \\
		\omega(\{n_e, n_i\}\to\{n_e, n_i+1\})&\approx(N_i-n_i)\langle f(\Lambda_i)\rangle.
	\end{align}
	\end{subequations}
	
	where $\langle f(\Lambda_{e,i}) \rangle$ represents formally the average input received by an individual. Using the approximated rates, the corresponding multivariate Fokker-Planck equation can be written as
	
	\begin{equation}
		\partial_t P(\rho_e,\rho_i,t)=\frac{\partial}{\partial{\rho_e}}\left(D^{(1)}_{e}P\right)+\frac{\partial}{\partial{\rho_i}}\left(D^{(1)}_{i}P\right)+\frac{1}{2}\frac{\partial^2}{\partial{\rho^2_e}}\left(D^{(2)}_{ee}P\right)+\frac{1}{2}\frac{\partial^2}{\partial{\rho^2_i}}\left(D^{(2)}_{ii}P\right),
	\end{equation}
	
	where we have omitted the dependency of the probability density function, $P$, in the right hand side to simplify the notation. The coefficients are

	\begin{subequations}	
	\begin{align}
		D^{(1)}_{e} =& -\mu_e\rho_e+(\alpha-\rho_e)\langle f(\Lambda_e)\rangle,\\
		D^{(1)}_{i} =& -\mu_i\rho_i+(1-\alpha-\rho_i)\langle f(\Lambda_i)\rangle,\\
		D^{(2)}_{ee} =& \frac{1}{N}\left[\mu_e\rho_e+(\alpha-\rho_e)\langle f(\Lambda_e)\rangle\right],\\
		D^{(2)}_{ii} =& \frac{1}{N}\left[\mu_i\rho_i+(1-\alpha-\rho_i)\langle f(\Lambda_i)\rangle\right],
	\end{align}
	\end{subequations}	
	where, the number of active nodes has been replaced by the density of active sites $\rho_{e,i} = n_{e,i} / N$. Finally, we can use the equivalence between the Fokker-Planck and Langevin equations to derive the following two stochastic equations under the Itô interpretation:
	
	\begin{subequations}
	\label{eqn:LangevinSI} 
	\begin{align}
		\dot\rho_e =& -\mu_e\rho_e + (\alpha-\rho_e)\langle f(\Lambda_e)\rangle + \frac{1}{\sqrt{N}} \sqrt{\mu_e \rho_e + (\alpha-\rho_e)\langle f(\Lambda_e)\rangle}\xi_e(t), \\
		\dot\rho_i =& -\mu_i \rho_i + (1-\alpha-\rho_i)\langle f(\Lambda_i)\rangle +\frac{1}{\sqrt{N}} \sqrt{\mu_i \rho_i + (1-\alpha-\rho_i)\langle f(\Lambda_i)\rangle}\xi_i(t),
	\end{align}
	\end{subequations}
	
	with $\xi_{e,i}(t)$ being uncorrelated, Gaussian white noises.

	\section{Mean-field analysis}
	Mean Field equations are obtained by averaging over noise the Langevin eqs. (\ref{eqn:LangevinSI}) and taking the approximation $\langle f(\Lambda_{e/i})\rangle\approx f(\langle\Lambda_{e/i}\rangle)$. 
	\subsection*{Symmetric version of the model}
	
	\begin{figure} 
		\includegraphics{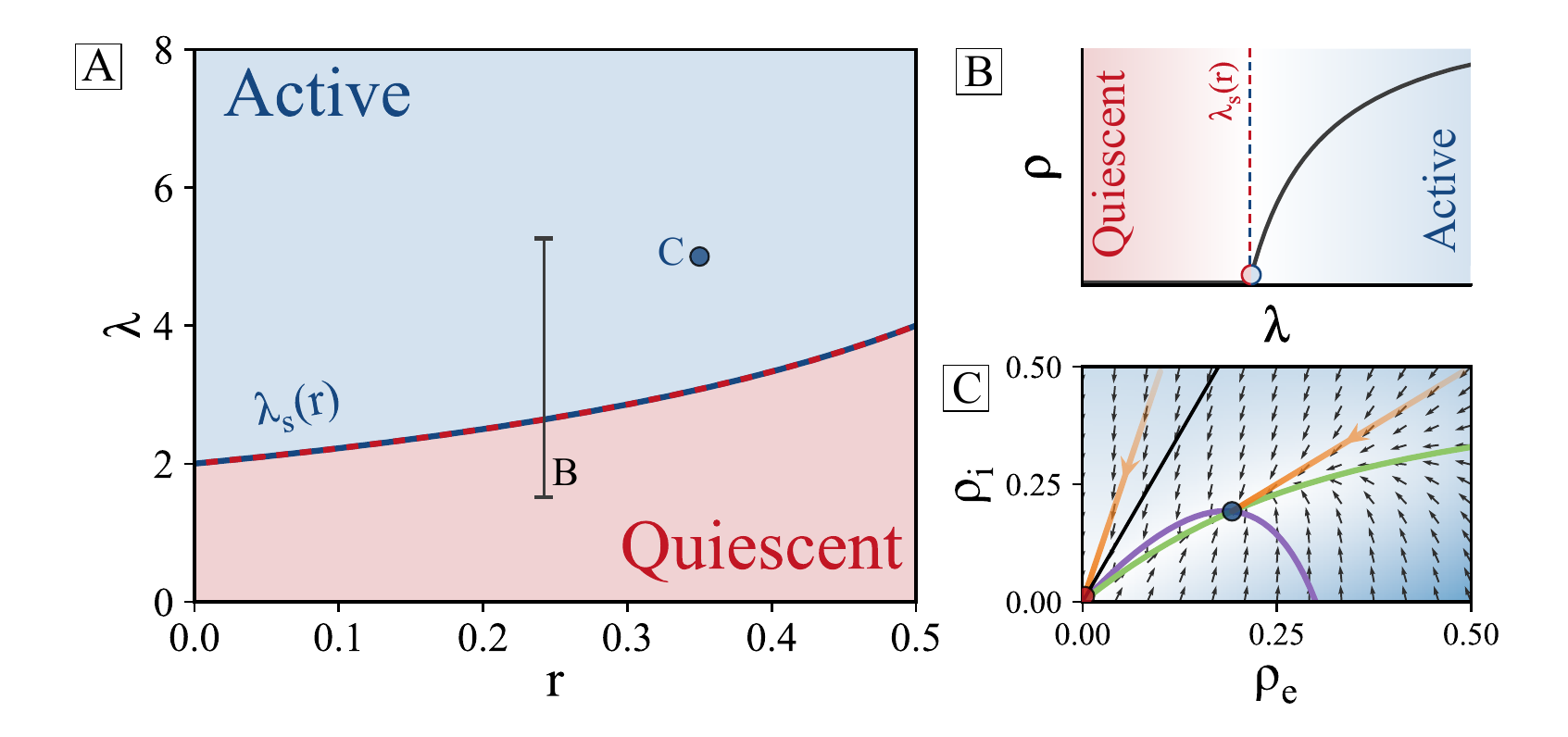}
		\caption{Phase diagram of the mean-field symmetric case for $\alpha=1/2$. (A) The symmetric case presents quiescent and active phases separated by a transcritical bifurcation. (B) Sketch of a vertical slice ($\rho$ vs $\lambda$ for fixed $r$) of the phase diagram shown in A. (C) Flow diagrams in the
		$\rho_e, \rho_i$ plane at the point marked
		and color-coded in panel (A).  The background color stands for the
		phase and its color intensity is proportional to the vector field
		module. The two nullclines are colored in green ($\dot{\rho}_e=0$)
		and purple color ($\dot{\rho}_i=0$), respectively and the black line
		($\rho_i=\rho_e/r$) separates zone 1 (inhibition dominated) from
		zone 2 (excitation dominated).  Characteristic trajectories starting in either of the zones are depicted as arrowed orange lines and the colored point stand for the stable
		steady state.}\label{fig:MFSym}
	\end{figure} 
	The deterministic mean-field equations for the EI-symmetric variant of the model ($r_e=r_i\equiv r$) are given by
	
	\begin{subequations}
	\label{eqn:MFSym}
	\begin{align}
		\dot \rho_e &= -\rho_e + (\alpha-\rho_e)\lambda f(\rho_e - r\rho_i)\, \\
		\dot \rho_i &= -\rho_i + (1-\alpha-\rho_i)\lambda f(\rho_e - r\rho_i).
	\end{align}
	\end{subequations}
	
	In order to obtain the fixed points ($\dot \rho_e=\dot \rho_i=0$) it is necessary to divide the phase plane into two different zones due to the presence of the non smooth function, $f$: an inhibition-dominated region (\emph{zone 1}), with $\rho_e<r\rho_i$, and an excitation-dominated one (\emph{zone 2}) where $\rho_e>r\rho_i$. Inside the former, the transfer function always returns a zero value, and, therefore, the only possible fixed point is the absorbing state $\rho_e=\rho_i=0$. Contrarily, inside zone 2 there is an active state, $\rho_e=\alpha\rho^*$ and $\rho_i=(1-\alpha)\rho^*$, with $\rho^*=1-1/\left(\lambda(\alpha - (1-\alpha)r)\right) $, as well as the absorbing state which is always present.
	
	The Jacobian of the system reads:
	
	\begin{widetext}
		\begin{equation}\label{eqn:SymmetricJacobian}
			J(\rho_e,\rho_i) = 
			\begin{pmatrix}
				-1 + \lambda \left[ (\alpha-\rho_e)f'(\rho_e-r\rho_i)-f(\rho_e-r\rho_i) \right] & -\lambda r(\alpha-\rho_e)f'(\rho_e-r\rho_i) \\
				\lambda(1-\alpha-\rho_i)f'(\rho_e-r\rho_i) & -1 -\lambda \left[ (1-\alpha-\rho_i)rf'(\rho_e-r\rho_i) +  f(\rho_e-r\rho_i) \right]
			\end{pmatrix},
		\end{equation}
	\end{widetext}
	
	with $f'(x)\equiv\theta(x)$ the Heaviside step function. In order to analyze the stability, one should study its eigenvalues evaluated in the different fixed points. When evaluating the Jacobian at the absorbing state, one notices that $f'(0)$ is not defined. Nonetheless, for any other point $f'$ is well behaved, therefore what one can do is to approach the origin from the two different zones defined above and evaluate its stability separately. Thus, approaching inside zone 1 ($\rho_e<r\rho_i$) gives
	\begin{equation}
		J(\vec{0}^\text{-}) = 
		\begin{pmatrix}
			-1  & 0 \\
			0 & -1  
		\end{pmatrix},
	\end{equation}
	which reveals that the absorbing state is a stable node for every value of $r$ and $\lambda$ inside zone 1. One can also compute the Henrici index, or amount of non normality, as defined in \ref{eqn:Henrici}
	\begin{equation}
		\mathcal{NN}(J(\vec{0}^\text{-})) = 0,
	\end{equation}
	which tell us that any initial trajectory inside zone 1 will directly relax to the quiescent state without showing any kind of amplified transient behavior (see trajectory in the inhibition-dominated region of Fig. \ref{fig:MFSym}C).
	
	Contrarily, coming close to the origin from zone 2 ($\rho_e>r\rho_i$) one gets
	\begin{equation}
		J(\vec{0}^\text{+}) = 
		\begin{pmatrix}
			-1+\lambda\alpha  & -\lambda r\alpha \\
			\lambda(1-\alpha) & -1 -\lambda r(1-\alpha)
		\end{pmatrix},
	\end{equation}
	with eigenvalues: $E_1=-1$, $E_2=-1+\lambda(\alpha-(1-\alpha)r)$. Then, it is straightforward to see that inside zone 2 the origin looses its stability at $\lambda=1/\left(\alpha-(1-\alpha)r\right)$, which for $\alpha=1/2$ gives $\lambda_s(r)=2/(1-r)$. In this case the Henrici index is given by:
	\begin{equation}
		\mathcal{NN}(J(\vec{0}^\text{+})) = \frac{\lambda}{2}(1+r),
	\end{equation}
	which can seem \emph{a priori} arbitrarily large if we increase $\lambda$. However, since the quiescent state is only stable below $\lambda_s(r)=2/(1-r)$ (for $\alpha=1/2$), the non normality of the quiescent phase is quite low for almost all $r$ and $\lambda$ values.
	
	Likewise, evaluating Eq. (\ref{eqn:SymmetricJacobian}) in the active fixed point gives the following eigenvalues: $E_1=-\lambda(\alpha-(1-\alpha)r)$, $E_2=1-\lambda(\alpha-(1-\alpha)r)$, evincing the gaining of stability of the active fixed point precisely at  $\lambda=1/\left(\alpha-(1-\alpha)r \right)$ resulting in a transcritical bifurcation which give raise to a continuous phase transition of the total activity $\rho$ as a function of $\lambda$ from quiescent to active (see Fig. \ref{fig:MFSym}A-B).

	\subsection*{Non-symmetric version of the model}
	
	The mean-field equations for the non-symmetric variant of the model, $r_i\neq r_e$, in the limit of $r_i=0$ are given by
	
	\begin{subequations}
	\begin{align}
		\dot \rho_e &= -\rho_e + (\alpha-\rho_e)\lambda f(\rho_e - r\rho_i), \\
		\dot \rho_i &= -\rho_i + (1-\alpha-\rho_i)\lambda \rho_e, 
	\end{align}
	\end{subequations}
	where $r_e\equiv r$. Dividing again into the two different zones defined above we obtain the following fixed points: $\rho^*_{e0}=\rho^*_{i0}=0$, which, as before, is present for both zones, and two active ones,
	$(\rho_{e-}^*,\rho_{i-}^*)$ and $(\rho_{e+}^*,\rho_{i+}^*)$, that are only present in zone 2, with
	
	\begin{subequations}
	\label{eqn:ActiveFixedPoints}
	\begin{align}
			\rho_{e\pm}^*=&\frac{1}{{2\lambda}}\left[\vphantom{\sqrt{x^2}}-2+\lambda(\alpha+r(1-\alpha)) \right. 
			\left.\pm\sqrt{\lambda}\sqrt{\lambda(\alpha-r(1-\alpha))^2-4 r(1-\alpha)}\right], \\
			\rho_{i\pm}^*=&\frac{1}{{2\sqrt{\lambda}(1+\alpha\lambda)r}}\left[\sqrt{\lambda}(r(1-\alpha)(1+2\alpha\lambda)-\alpha)\right. 
			\left.\pm\sqrt{\lambda(\alpha-r(1-\alpha))^2-4r(1-\alpha)}\right].
	\end{align}
	\end{subequations}

	The Jacobian reads
	
	\begin{equation}\label{eqn:NonSymmetricJacobian}
		J(\rho_e,\rho_i) = 
		\begin{pmatrix}
			-1 + \lambda \left[ (\alpha-\rho_e)f'(\rho_e-r\rho_i)-f(\rho_e-r\rho_i) \right] & -\lambda r(\alpha-\rho_e)f'(\rho_e-r\rho_i) \\
			\lambda(1-\alpha-\rho_i) & -1 -\lambda \rho_e 
		\end{pmatrix}.
	\end{equation}
	
	Similarly to the symmetric case, we study the stability of the origin in the two different zones. Then, approaching the origin from zone 1 gives
	\begin{equation}
		J(\vec{0}^\text{-}) = 
		\begin{pmatrix}
			-1  & 0 \\
			\lambda(1-\alpha) & -1  
		\end{pmatrix},
	\end{equation}
	which is a non-normal matrix with one degenerate eigenvalue $E_{1,2}=-1$ and only one eigenvector $\vec v_1=(0,1)$. In the language of linear stability analysis, this peculiar case is commonly known as a \emph{defective}, or \emph{degenerate}, node \cite{Strogatz}, which in our case turns out to be stable for any value of $r$ and $\lambda$ since the trace of the Jacobian is always negative. Computing the Henrici index gives:
	\begin{equation}
		\mathcal{NN}(J(\vec{0}^\text{-}))=\lambda/2
	\end{equation}
	which can be arbitrarily large since the quiescent state inside zone 1 is stable for all $\lambda$ values and therefore can describe large transient behaviours as shown in Fig.2G of the main text.
	
\begin{figure} 
		\includegraphics[width=0.9\textwidth]{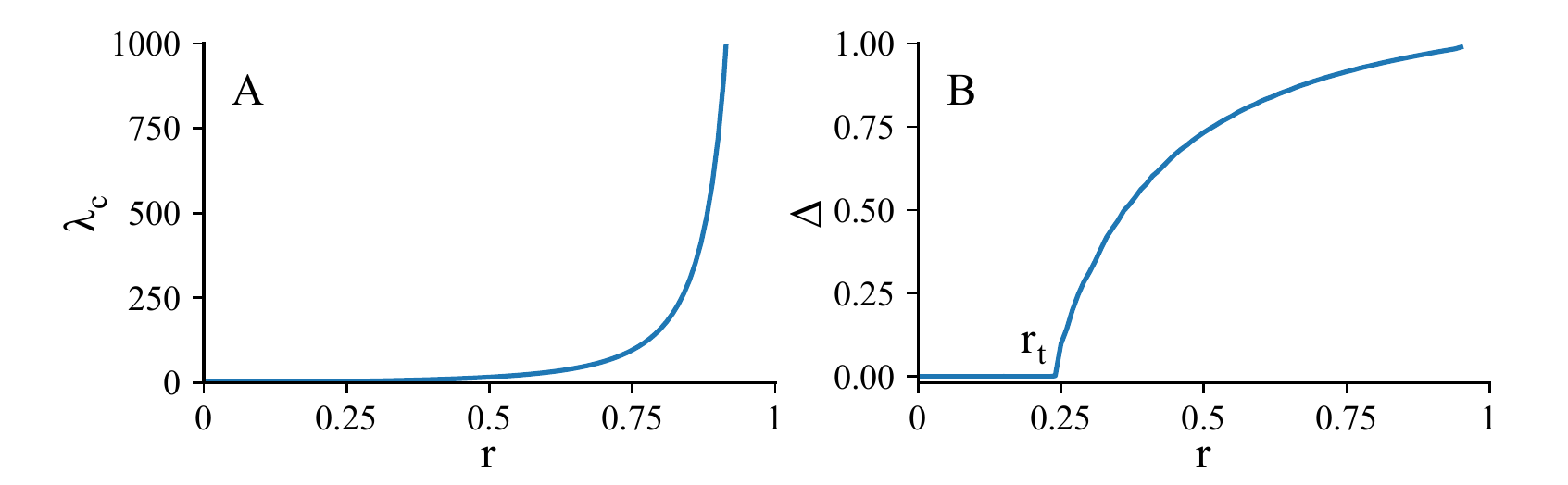}
		\caption{Features of the non-symmetric version of the model. (A) Critical point,
			$\lambda_c$, as a function of the strength of
			inhibition, $r$, showing the big shift in the critical point for large values of $r$. (B) Width of the discontinuity,
			$\Delta$, as a function of $r$. Results obtained solving numerically the mean-field equations for $\alpha=1/2$.}\label{fig:MFproperties}
	\end{figure}

	On the other hand, approaching the absorbing state from zone 2 we have
	\begin{equation}
		J(\vec{0}^\text{+}) = 
		\begin{pmatrix}
			-1+\lambda\alpha  & -\lambda r\alpha \\
			\lambda(1-\alpha) & -1 
		\end{pmatrix},
	\end{equation}
	with $E_{1/2}=-1+\frac{1}{2}\lambda\left(\alpha\pm\sqrt{\alpha^2-4\alpha r(1-\alpha)}\right)$ as eigenvalues. Then, local stability of the absorbing state is guaranteed for
	
	\begin{equation}
		\begin{cases} 
			\lambda<\frac{2}{\alpha+\sqrt{\alpha^2-4r\alpha(1-\alpha)}} & r\leq \frac{\alpha}{4(1-\alpha)}, \\
			\lambda<\frac{2}{\alpha} & r> \frac{\alpha}{4(1-\alpha)}, \\ 
		\end{cases}
	\end{equation}
	that in the case of $\alpha=1/2$ reduces to $\lambda_1(r)$ and $\lambda_2(r)$ of the main text. Again calculating the Henrici index gives:
	\begin{equation}
		\mathcal{NN}(J(\vec{0}^\text{+})) =
		\begin{cases} 
			\lambda(1-\alpha(1-r)) & r\leq \frac{\alpha}{4(1-\alpha)}, \\
			\lambda\sqrt{\alpha^2+(1-\alpha(1+r))^2} & r> \frac{\alpha}{4(1-\alpha)}, \\ 
		\end{cases}
	\end{equation}
	which can also display a large non normality for certain values of $\lambda$ and $r$. More concretely and for $\alpha=1/2$ the whole \emph{excitable} phase (Fig.2A of the main text) can exhibit large values of non normality by making $\lambda$ large , since now the global stability of this excitable quiescent phase is bounded by $\lambda_3(r)=8r/(r-1)^2$ (see below), which is indeed much bigger ---for the same $r$ values--- than its counterpart in the symmetric version of the model (see Fig.\ref{fig:MFproperties}A).
	
	Finally, in order to see the stability of the two active points $(\rho_{e\pm}^*, \rho_{i\pm}^*)$ one may proceed as before, evaluating the eigenvalues of Jacobian in eq. (\ref{eqn:NonSymmetricJacobian}) in those fixed points. However, given the rather involved form of the fixed points, closed analytical expressions for stability conditions are difficult to find in this way. Therefore, initially detailed numerical evaluations were performed, finding that the point $(\rho_{e+}^*, \rho_{i+}^*)$, is always stable whereas the other, $(\rho_{e-}^*, \rho_{i-}^*)$ is always unstable. Thus, in order to find the stability conditions for $(\rho_{e+}^*, \rho_{i+}^*)$, one just needs to find the region of parameters $r$ and $\lambda$ where this fixed point appears, i.e., the region where: the density of excitatory and inhibitory is real and non-negative, $\rho_{e/i+}^*\in\mathbb{R}^+_0$; and additionally, where this densities obey the consistency condition $\rho_{e+}^*>r\rho_{i+}^*$, since they were obtained in zone 2. All this, together with $\alpha=1/2$ for simplicity, gives
	\begin{equation}
		\begin{cases}
			\lambda>\frac{4}{1+\sqrt{1-4r}} & r\leq -2+\sqrt{5}, \\
			\lambda>\frac{8r}{(r-1)^2} & r> -2+\sqrt{5}, \\ 
		\end{cases}
	\end{equation}
	giving $\lambda_1(r)$ and $\lambda_3(r)$ of the main text.

	Fig. \ref{fig:MFproperties}A illustrates the dependence of the transition point $\lambda_c$ 
	on the strength of the inhibition, $r$. Observe that as $r
	\rightarrow 1$, i.e. as inhibition tends to completely block
	excitatory activations, the value of the activation rate required to create an
	active phase diverges. Furthermore, Fig. \ref{fig:MFproperties}B illustrates the fact that as $r$ is increased, there is a certain value $r_t$ at which the transition
	becomes discontinuous and the associated ``jump'', $\Delta=\rho_{e+}(\lambda_c(r))$,  grows with $r$.
	
	The rationale for the emergence of a discontinuous transition in the EI non-symmetric version of the model is as follows: for small values of $r$, what matters is the competition between the activation and inactivation rate, i.e., the branching ratio $m=\lambda/\mu$, much as in the standard CP. As the branching ratio is increased, activation overcomes inactivation, and since the inhibition feedback $r$ is small, both $\rho_e$ and $\rho_i$ can maintain an arbitrarily low value, leading to a continuous transition. Mathematically, this is explained by the fact that the fixed points exchange stability at the same value of $\lambda$ (Fig.2A-B of main text). On the other hand, once $r$ increases over $r_t=-2+\sqrt{5}$, excitatory units cannot maintain an arbitrarily low value due to the effect of inhibition, despite having a sufficiently large branching ratio. Once the excitatory units turn off, the inhibition will inevitably follow the same fate.  Upon further increasing the branching ratio,  one eventually arrives at a certain point at which activation effect overcomes both inactivation and inhibition feedback, so the system can maintain a value of $\rho_e>0$ which is not arbitrarily low, leading to a discontinuous transition.

	Observe that, on the other hand, in the EI symmetric version the transition is always continuous because both excitatory and inhibitory units are equally inhibited. Therefore, once the $\lambda$ effect overcomes $\mu$ and $r$ (for any value of $\lambda$ and $r$), we can have an arbitrarily low value for both $\rho_e$ and $\rho_i$ at the same $\lambda$ value, thus leading to a continuous transition.

	\section{Results beyond mean-field \label{sec:lattice}}

	\subsection{Jensen's force and the asynchronous phase}

	Let us analyze the full system of deterministic equations, assuming the non-symmetric case ($r_i=0$, $r_e\equiv r$), although the deduction for the symmetric case is similar,
		
	\begin{subequations}
	\label{eqn:deterministic-full} 
	\begin{align}
		\dot\rho_e =& -\mu_e\rho_e + (\alpha-\rho_e)\langle f(\Lambda)\rangle, \\
		\dot\rho_i =& -\mu_i \rho_i + \lambda (1-\alpha-\rho_i) \rho_e.
	\end{align}
	\end{subequations}
	
	In the mean field analysis we approximated $\langle f(\Lambda)\rangle$ by $ f(\langle\Lambda\rangle )$ where $f(\Lambda)\equiv\max(0, \Lambda)$, this is exact for fully connected networks and a good approximation for sparse networks for large values of $\langle \Lambda \rangle$ since $f$ is linear in that limit. However, for small $\langle \Lambda \rangle$, fluctuations (due to sparsity) in the input received by each node can produce a negative value leading to $f(\Lambda < 0) = 0$ which averaging over all possible inputs lead to $\langle f(\Lambda)\rangle\geq f(\langle \Lambda\rangle)$. This is known as the Jensen's inequality and is valid for any convex function. Therefore, input fluctuations translate into a stochastic force, $F(\Lambda)=\langle f(\Lambda)\rangle - f(\langle \Lambda\rangle)$ (termed Jensen's forced in \cite{Buendia-Jensen}) that can render the quiescent state (stable in the mean field case) unstable, generating a low-activity phase in sparse networks.

	A simple way to go beyond the mean-field approximation is to follow the procedure of Buend\'ia et al. \cite{Buendia-Jensen}, assuming that the system is annealed, i.e., connections are constantly reshuffling: every time the interactions of a node are evaluated, $k$ neighbours are randomly selected, of which $k_e=\alpha k$ are excitatory and $k_i=(1-\alpha)k$ are inhibitory. 
	
	Let us derive the equations for this approach. Since there is a fraction of $\rho_e\in[0,\alpha]$ active excitatory nodes, the probability of picking an active node when an excitatory unit is selected at random is given by $\rho_e/\alpha$. Then, the probability of obtaining $j$ active nodes when taking $k_e=k\alpha$ random excitatory nodes is given by the binomial probability
	
	\begin{equation}
	b(j,k_e;\rho_e) = \binom{k_e}{j} \left( \frac{\rho_e}{\alpha} \right)^j \left(1-\frac{\rho_e}{\alpha}\right)^{k_e-j}.
	\end{equation}
	
	The same argument can be done with the inhibitory population, so the probability of taking $l$ out of $k_i$ active inhibitory nodes is also given by a binomial, $b(l,k_i;\rho_i)$. Now, notice that if one knows the number of both active excitatory and inhibitory connections, one knows the input to the target node. Then, the probability of observing a certain input can be determined by the product of both binomials,
	
		\begin{equation}
		p_{jl}(\rho_e,\rho_i) = \binom{k_e}{j} \binom{k_i}{l}  \left( \frac{\rho_e}{\alpha} \right)^j \left(\frac{\rho_i}{1-\alpha}\right)^l     \left(1-\frac{\rho_e}{\alpha}\right)^{k_e-j} \left( 1- \frac{\rho_i}{1-\alpha}   \right) ^{k_i-l}. 
		\end{equation}	
		
	Finally, we can use this probability in order to evaluate the average of the input function, 
	
	\begin{equation}
	\langle f(\Lambda) \rangle = \frac{1}{k} \sum_{j,l=0,0} ^{j,l=k_e, k_i} f\left( \lambda(j-r l) \right) p_{jl}(\rho_e,\rho_i),
	\label{eq-exact-f-average}
	\end{equation}

	For low activity, the average can be exactly evaluated up to first order in $\rho_e,\rho_i$. Note that any combination of $j+l>1$ will be of order larger than 2, which means that we only need to consider the contributions $j,l=1,0$ and $j,l=0,1$ to the sum. As long as $r\in[0,1]$, the result of the transfer function does not depend on $r$ itself, meaning that the sum can be approximated to 
	
	\begin{equation}
	\langle f(\Lambda) \rangle \simeq  \frac{\lambda k_e \rho_e}{\alpha}.
	\end{equation}
	
	If this result is plugged into eqs. (\ref{eqn:deterministic-full}), it is possible to find the fixed points and their stability for the annealed system in the limit of low activity. The Jacobian at the quiescent state is given by 
	
		\begin{equation}
			J(0,0) = 
			\begin{pmatrix}
				-1 + \alpha \lambda & 0 \\
				(1-\alpha)\lambda   & -1 
			\end{pmatrix}.
		\end{equation}	
	
	We then find that the quiescent state loses stability at the value $\lambda=1/\alpha$, no matter the value of the other parameters. Notice that this value is the same as the one obtained for the excitatory-only case, $r=0$. Moreover, the Jacobian displays a non-normal form, meaning that this novel low-activity phase, that we call Asynchronous (AS) Phase (since it displays all the properties of the cortical asynchronous state as shown in the main text), is always excitable (see Sec.\ref{sec:reactivity} for a detailed explanation of the relation between non normality and excitability).
	
	In the general case, the Jacobian is given by:
	
	\begin{equation}\label{eqn:AnnealedJacobian}
		J(\rho_e,\rho_i) = 
		\begin{pmatrix}
			-1 - \langle f(\Lambda) \rangle + (\alpha-\rho_e)\partial_e \langle f(\Lambda) \rangle & (\alpha-\rho_e)\partial_i \langle f(\Lambda) \rangle  \\
			(1 - \alpha - \rho_i)\lambda    &  -1 - \lambda \rho_e
		\end{pmatrix},
	\end{equation}	
	
	with the derivatives of $\langle f(\Lambda) \rangle $ being
	
	\begin{align}	\label{eq-f-derivatives}
		\partial_e \langle f(\Lambda) \rangle &= \frac{1}{k} \sum_{j,l} f\left( \lambda(j-r l) \right) \binom{k_e}{j} \binom{k_i}{l} \left( \frac{\rho_e}{\alpha} \right)^{j-1} \left(\frac{\rho_i}{1-\alpha}\right)^l     \left(1-\frac{\rho_e}{\alpha}\right)^{k_e-j-1} \left( 1- \frac{\rho_i}{1-\alpha}   \right) ^{k_i-l} \left( j - \frac{k_e \rho_e}{\alpha} \right), \\
		\partial_i \langle f(\Lambda) \rangle &= \frac{1}{k} \sum_{j,l} f\left( \lambda(j-r l) \right) \binom{k_e}{j} \binom{k_i}{l} \left( \frac{\rho_e}{\alpha} \right)^j \left(\frac{\rho_i}{1-\alpha}\right)^{l-1}     \left(1-\frac{\rho_e}{\alpha}\right)^{k_e-j} \left( 1- \frac{\rho_i}{1-\alpha}   \right) ^{k_i-l-1} \left( l - \frac{k_i \rho_i}{1-\alpha} \right).
	\end{align}
	
	Thus, one can integrate the differential equations \ref{eqn:deterministic-full} plugging \ref{eq-exact-f-average} until convergence to the stable state is reached, and this value can be fed into the Jacobian. Numerical diagonalization of this Jacobian yields the eigenvalues corresponding to our point. We applied this procedure in a $300\times300$ grid. The resulting plot can be seen in Fig. \ref{fig:Annealed} showing the phase diagram and real and imaginary parts of the largest eigenvalue for both the asymmetric and symmetric cases. The phenomenology resembles very much that of their mean-field counterpart with some key differences:
	\begin{figure} 
		\includegraphics[width=0.87\textwidth]{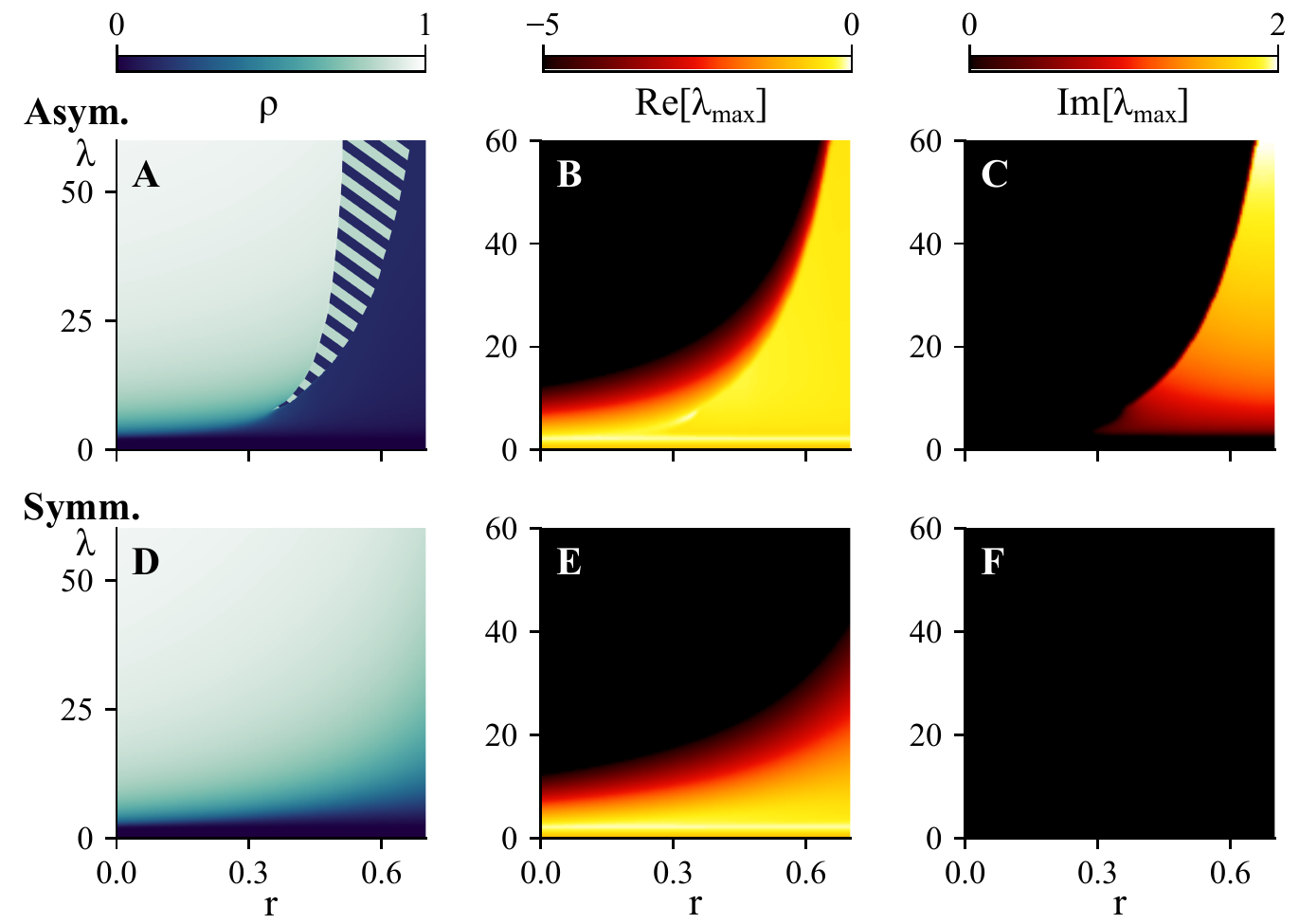}
		\caption{EI-CP on the annealed network. (A,D) Stationary activity for the asymmetric (A) and symmetric (D) versions of the model where stripes indicates the bistability region between the usual active phase (up state) and asynchronous phases. (B-F) Real (B,E) and imaginary (C,F) parts of the maximum eigenvalue of the Jacobian in eq.(\ref{eqn:AnnealedJacobian}) evaluated at the stationary state of the dynamics for the (B,C) asymmetric variant of the model, where in the bistable region the up state is shown and (E,F) symmetric variant.} \label{fig:Annealed}
	\end{figure}

	\begin{enumerate}
	\item The quiescent phase loses stability always at $\lambda = 1/\alpha$, as predicted by our low-activity expansion before. This can be clearly seen in Fig.\ref{fig:Annealed}.B,E, where the largest eigenvalue of the stationary state of the dynamics is depicted and so values close to zero represent transitions. This transition is the typical quiescent to active for the symmetric case, whereas for the asymmetric case it becomes a quiescent to AS transition (see Fig.\ref{fig:Annealed}.A,D).
	\item In the asymmetric version of the model, the line separating quiescent and active phases now splits: the system has a critical quiescent-AS transition, and then an AS-active separation, which is no longer critical and can be either continuous or discontinuous depending on the specific $r$ and $\lambda$ values.
	\item The Hopf bifurcation of the mean-field diagram of the asymmetric case does disappear. It translates into a smooth line separating two regions with different degrees of excitability, quantified by the value of the eigenvalues complex part (see Fig.\ref{fig:Annealed}C).
	\item Bistability regions of the mean field diagram for the asymmetric case between a locally or globally stable quiescent state and an active one translate into a bistable region between a down state (AS phase) and an up state (active phase) (see Sec.\ref{sec:Bistability}). 
	\end{enumerate}

	\subsection{Bistability in sparse networks} \label{sec:Bistability}
\begin{figure} 
		\includegraphics{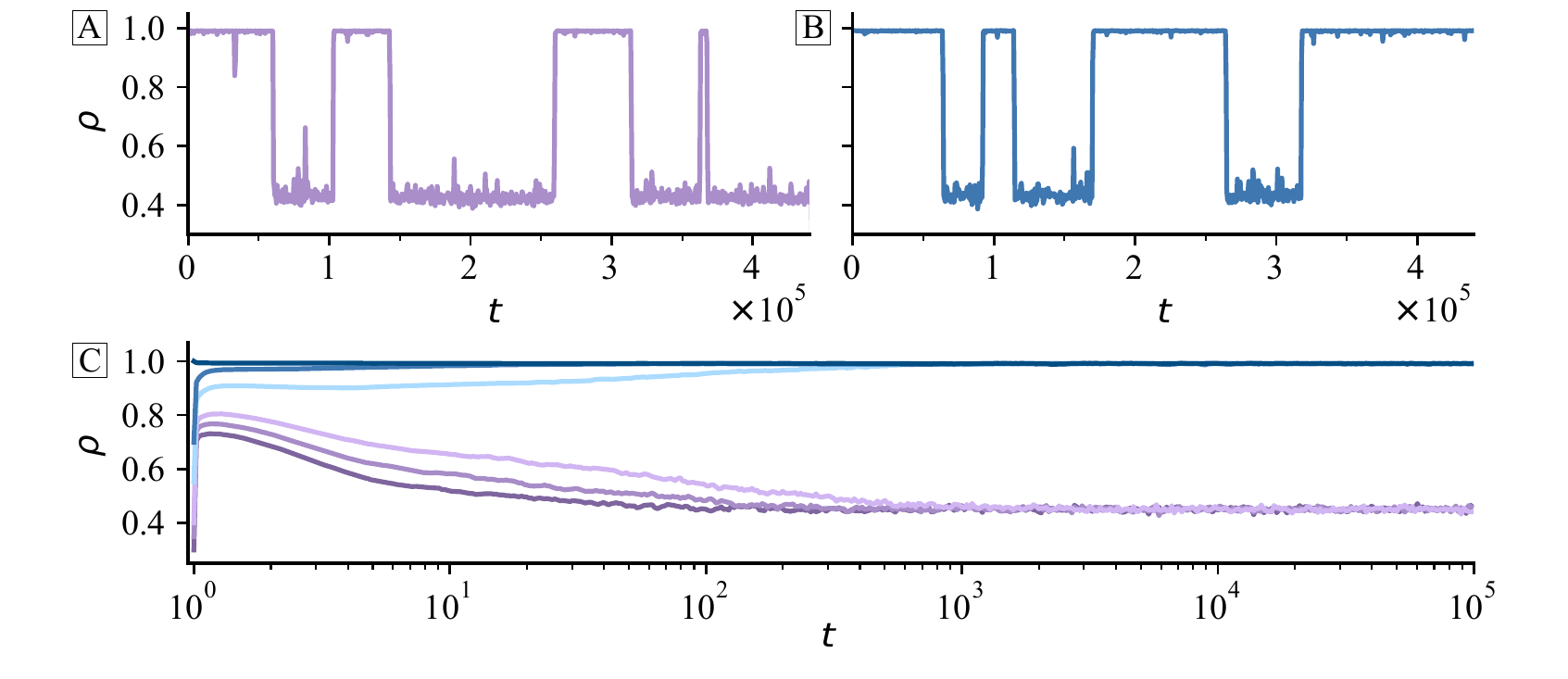}
		\caption{Demonstration of the existence of bistability in the vicinity of the AS phase to active phase transition with $r=0.7$. (A-B) Total activity of a single realization, $\rho$, as a function of time, $t$, in a highly noisy system, $N=1024$, for: (A) a down-dominated state,  $\lambda=700$; (B) an up-dominated state, $\lambda=720$. (C) Total activity of a single realization $\rho$ vs $t$ for $\lambda =750$ and different initial conditions in a system with much more moderate noise $N=10000$, resulting into two clearly different steady states, the down one (lillac colors) and the up one (blue colors). Results obtained in the 2D lattice ($\alpha=0.5$, $K=8$).} \label{fig:Bistability}
	\end{figure}

	It is known that during deep sleep or under anesthesia, the cerebral cortex exhibits bistability, with an alternation between high and low levels of neural activity, called up and down states respectively \cite{Steriade, Destexhe}. Our model is able to reproduce some form of bistability for an interval of $\lambda$ that grows with increasing $r$. Fig.4A,C of main text show this coexistence between the novel asynchronous state (down state) and the fully active (up state) in the 2D lattice and annealed cases. The existence of bistability in the 2D lattice is further investigated in Fig.\ref{fig:Bistability} where it can be easily visualized that for a small lattice (highly noisy), the system alternates between the down and up states for values of $\lambda\in[675, 775]$ approximately. Besides, the same figure also shows how different initial conditions result in two different steady states, the AS phase (lilac colors) and the active one (blue colors) which is another footprint of bistability. Those results are obtained for $r=0.7$. Similar analysis have also been carried out for smaller values of $r$ resulting in a smaller size of the bistability region, in accordance with the qualitative picture given by the annealed model phase diagram.	
	
	\subsection{Stochastic Amplification of Fluctuations}
\begin{figure} 
		\includegraphics[width=0.9\textwidth]{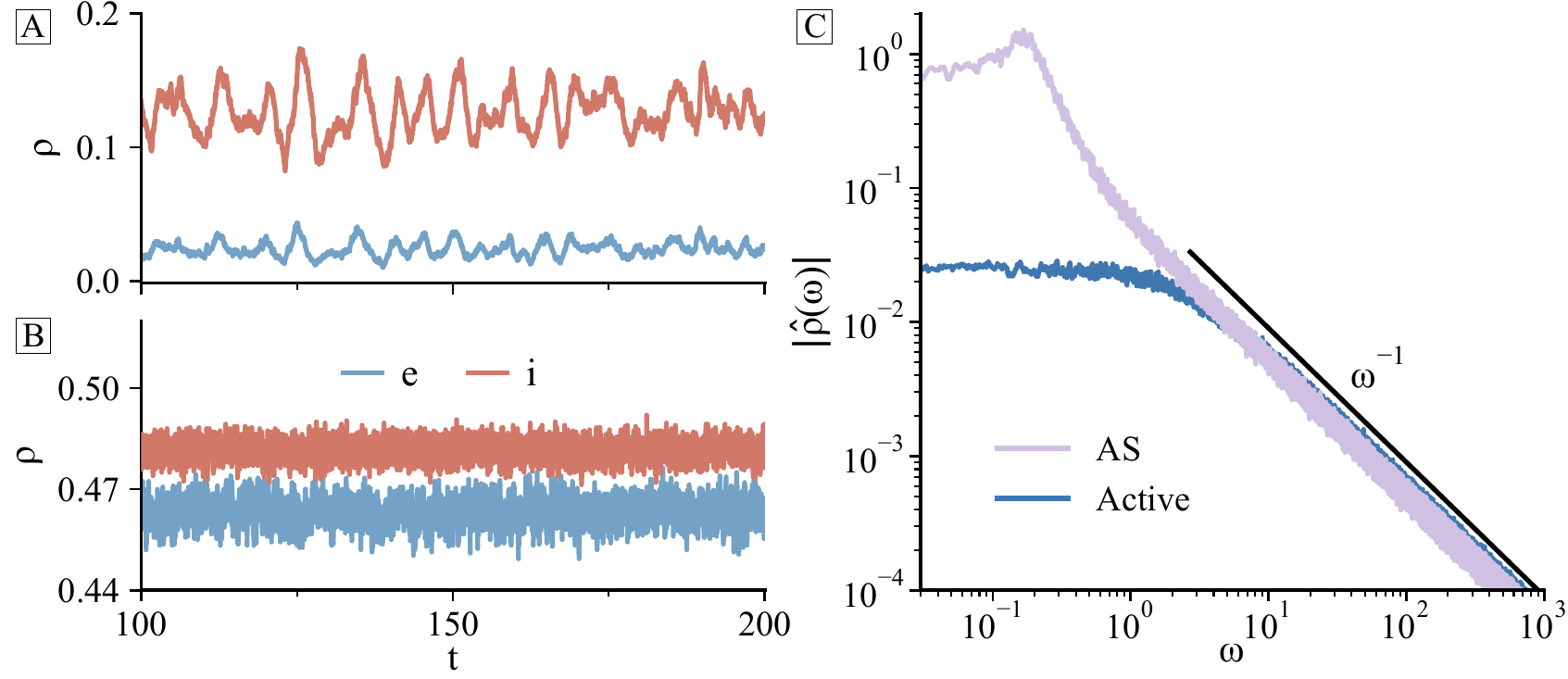}
		\caption{Illustration of the SAF phenomena. Panels (A) and (B) show the temporal evolution of excitatory (blue) and inhibitory (red) activity in the AS phase ($r=0.5$ and $\lambda=14$) and active phase ($r=0.5$ and $\lambda=57$) respectively. Panel (C) shows the absolute value of the Fourier transform, $|\hat\rho(\omega)|$ (averaged over 10 runs and smoothed out with a moving-averaged) as a function of frequency $\omega$ for the total activity, $\rho=\rho_e+\rho_i$, in the AS and active phases.  Results obtained in the annealed network for $\alpha=1/2$ and $K=30$. } \label{fig:SAF}
	\end{figure}

	Stochastic Amplification of Fluctuations (SAF) is a phenomena by which a system possessing a stable fixed point with complex eigenvalues (stable spiral) gets its relaxation trajectory to the stationary state frustrated by the noise, forcing it to quasi-oscillate. Interestingly enough, the time-series of activity of our model inside the AS phase exhibit a similar behavior, shown in Fig.\ref{fig:SAF}, where a peak can be found in the absolute value of the Fourier transform of activity, $|\hat \rho(\omega)|$; whereas the fully active case does not exhibit any oscillations whatsoever. Besides, both activities exhibit a perfect $1/\omega^2$ background noise in the power spectrum which is nothing but $|\hat \rho(\omega)|^2$.

	\subsection{Differences between Symmetric and Asymmetric cases \label{sec:differences}}
\begin{figure}
		\includegraphics{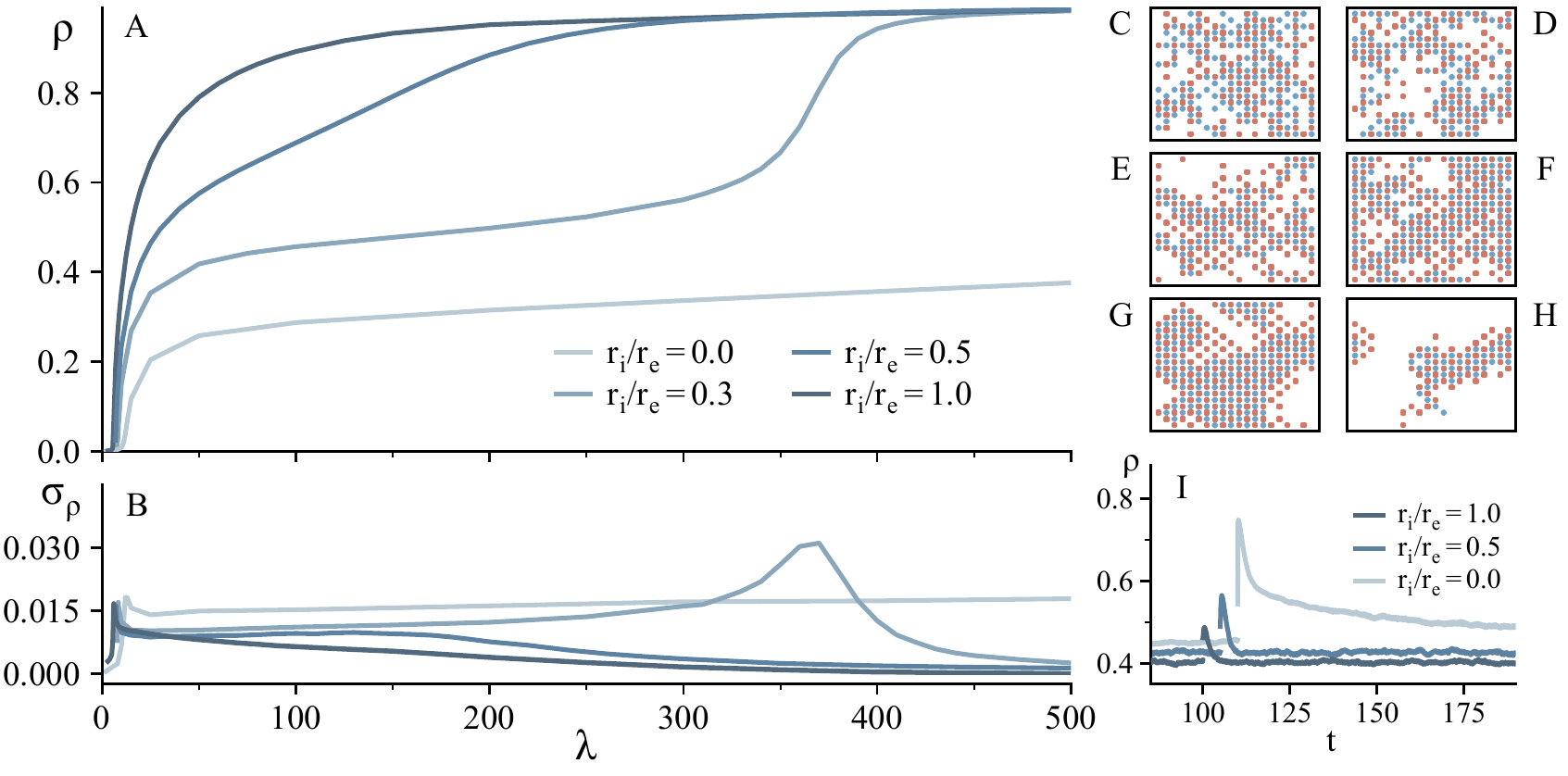}
		\caption{Differences between symmetric and asymmetric cases in the 2D lattice ($\alpha=0.5$, $K=8$) with $r_e=0.7$. Panels (A) and (B) show the activity, $\rho$, and its standard deviation, $\sigma_\rho$, respectively, as a function of the activation rate $\lambda$ for different fractions of $r_i/r_e$ going from the completely asymmetric case (light blue), $r_i/r_e=0$, to the completely symmetric model (dark blue), $r_i/r_e=1$ . The diagram is zoomed in the region $\lambda\in[0,500]$ to highlight the differences. Panels (C-H) show two different snapshots of the activity of a 2D lattice with $L=20$ for: $r_i/r_e=1$, $\lambda=20$ (C-D); $r_i/r_e=0.5$, $\lambda=80$ (E-F); and $r_i/r_e=0$, $\lambda=750$ (G-H). Panel I shows the temporal evolution of activity in a stimulation experiment, where a fraction $\Delta_e=0.15$ of excitatory nodes is instantaneously activated for: the totally symmetric case, $r_i/r_e=1$, with $\lambda=13$; the fully antisymmetric case, $r_i/r_e=0$ with $\lambda=750$; and an intermediate case, $r_i/r_e=0.5$ with $\lambda=13$. }\label{fig:Differences}
	\end{figure}

	In the main text we have studied thoroughly the asymmetric case because the symmetric case exhibits a similar behavior to the low-$r$ regime of the asymmetric one, i.e. even though the model still present some features of the asynchronous state for some $\lambda$ values, the excitability against perturbations, low activity regime for a wide region of $\lambda$ and quasi-oscillatory behaviour are lost. Indeed, Figs.\ref{fig:Differences}A,I illustrate how the low self-sustained activity and excitability are gradually lost as the symmetry in the coupling between excitatory and inhibitory neurons is increased. Both changes can be understood together by looking at the activity in the network (Fig.\ref{fig:Differences}C-H and Supplementary Videos 1-3), where one can see that by symmetrizing the model, the system changes from a highly fluctuant phase to a more static one. In the former, activity is concentrated in large clusters that travel around the network, breaking and recombining with others (Fig.\ref{fig:Differences}G,H and Suppementary Video 1); while the later is characterized by a less and less fluctuant and clusterized network, where activity cannot travel across the network (Fig.\ref{fig:Differences}C-F and Suppementary Videos 2,3). 

	\section{Methods\label{sec:methods}}

	\subsection{Scaling behaviour and critical exponents}
\begin{figure}
		\includegraphics{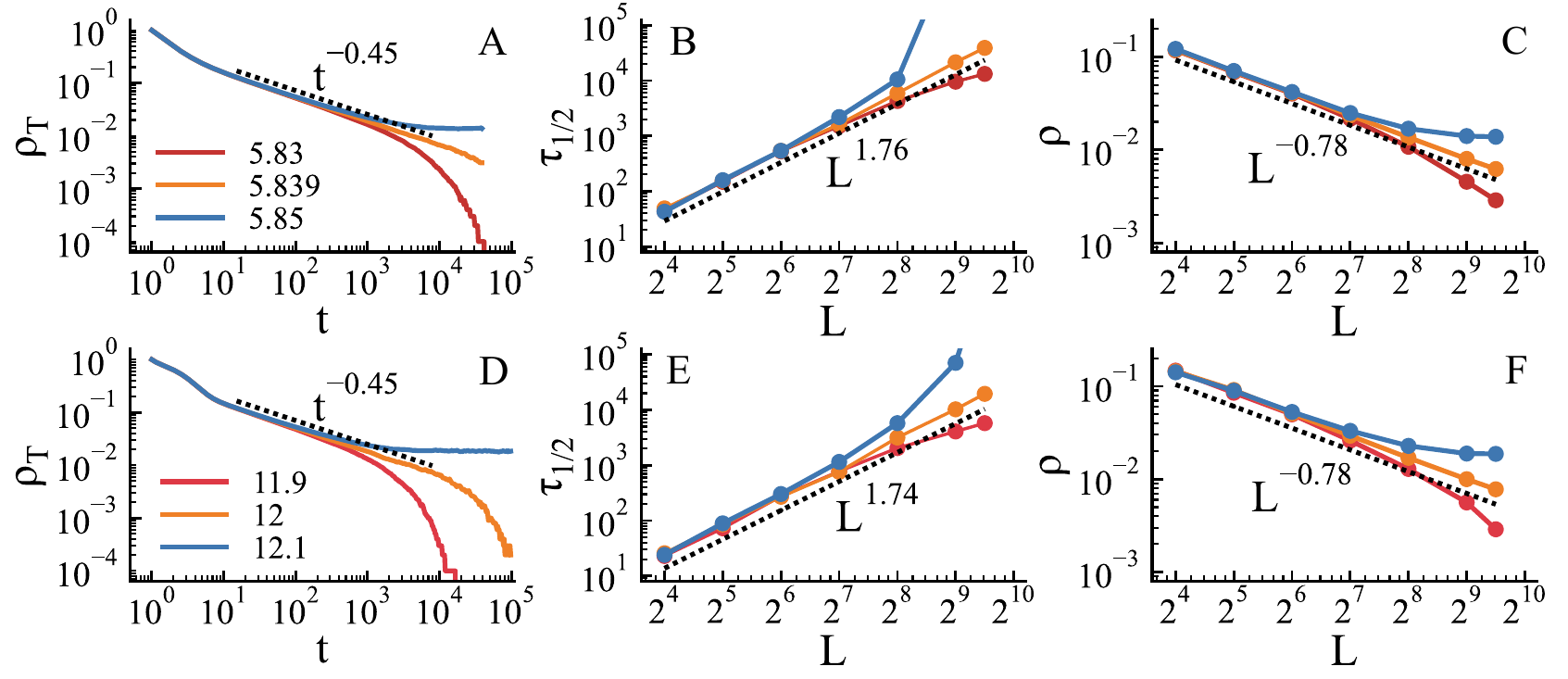}
		\caption{Finite size scaling for the symmetric (upper row) and asymmetric (lower row) cases of our model for values of $\lambda$ close to the critical point detailed in the legend. A,D: time decay of the activity averaged over all trials, $\rho_T$. Exponent value obtained in the range [15, 2000]. B,E: Characteristic time to reach the quiescent state, $\tau_{1/2}$ as a function of the linear size of the system $L$. C,F: Quasistationary value of the activity averaged over surviving trials, $\rho$ as a function of $L$. All of the analyses have been performed for a minimum of 300 runs in a 2D Lattice using the Gillespie algorithm.}\label{fig:FSS}
	\end{figure}

	In order to elucidate the nature of the phase transition between quiescent and active phases in the presence of inhibition we perform standard finite size scaling analyses \cite{Marro, Henkel} in a 2D lattice (that can display richer behaviour compared to higher-dimensional systems as networks), where we measure:
	\begin{enumerate}
		\item The quasistationary value of the activity averaged only for surviving trials, $\rho$, which scales as $\rho(L)\propto L^{-\beta/\nu_\perp}$ right at the critical point (Fig.\ref{fig:FSS}C,F). This density is estimated by computing it as a function of time and averaging its value where the density saturates, becoming a constant value.
		\item The characteristic time to reach the quiescent state, $\tau_{1/2}$, defined as the time required for the survival probability to decay to one half, which scales as $\tau_{1/2}\propto L^{\nu_{\parallel}/\nu_{\perp}}$ at the critical point (Fig.\ref{fig:FSS}B,E). Since one can only evaluate a finite number of runs, it may happen that some of the sizes do not arrive exactly to one  half of the probability. In that case we take the closest number to such half (the maximum difference in our case being of 0.02).
		\item The time-decay of the activity averaged over all trials (including those that have reached the absorbing state), $\rho_T$, that is expected to scale at the critical point as $\rho_T(L,t)\propto t^{-\delta}$ for times small compared with $\tau_{1/2}$ (Fig.\ref{fig:FSS}A,D).
	\end{enumerate}

	The resulting exponents obtained for the asymmetric (resp. symmetric) case are: $\beta/\nu_\perp\approx 0.78$ ($0.78$), $\nu_{\parallel}/\nu_{\perp}\approx 1.74$ ($1.76$), $\delta\approx 0.45$ ($0.45$), all of them in close agreement with their expected values in the two-dimensional DP class \cite{CP-Exponents}. We thus conclude that inhibition does not affect the nature of the phase transition neither for the symmetric case, nor for the asymmetric one. In this way, we also demonstrate that the transition between a quiescent state and the novel low-activity phase (asynchronous irregular state) still belongs as well to the DP universality class, in agreement with the Janssen-Grassberger conjecture, despite the special properties of the asynchronous irregular phase.

	\subsection{Avalanches}
     Avalanches are obtained by integrating Langevin equations \ref{eqn:LangevinSI}a-b --interpreted in the Itô sense-- with an Euler-Mayorama algorithm, which are computationally cheaper to simulate in comparison with performing Gillespie simulations of the full system in mean-field. We consider that an avalanche ends once the activity crosses below a certain (low) threshold. The value of the threshold is used also as initial condition, mimicking single-seed experiments in lattices. From these avalanches, one can compute several observables, such as the avalanche shape. This is obtained by selecting all avalanches with a fixed duration $T$, and averaging the activity $\rho(t/T)$ during the chosen avalanche duration.  
     Since the number of avalanches is finite, we have to defined a criterion to declare if two different avalanches had the same duration, $T$, or not. Thus, we perform a binning with relative accuracy of $0.1$ in real space and so we average together avalanches with $T\in [t^{(n)}, t^{(n+1)}]$ being $t^{(n+1)}=1.1t^{(n)}$ and $t^{(0)}=1$ (which is essentially a equidistant logarithmic binning of $\log_{10}(1.1)$). Other criterions such a equidistant real space binning of $1$ had also been taken into account and the qualitative picture did not change at all. Parameters used for the simulation are:  time-step, $dt=10^{-4}$, initial condition, $\rho_e=10^{-8}$, $\rho_i=0$, gaussian noises amplitudes, $\sigma_{e,i}=10^{-4}$, and a threshold of $10^{-8}$. The necessary amount of  total avalanches to display a clear collapse of the avalanche shape --shown in Fig.3 of the main text-- is of order $10^7$ for each of the different values of $r$ and $\lambda$.

	\subsection{Asynchronous-state-related observables}
	
	\emph{Coefficient of variation (CV).} It is defined as the ratio of the standard deviation to the mean of the interspike intervals (ISI) ---i.e. periods of un-interrupted silence for a given neuron/node.
	\begin{equation}
		\text{CV}=\frac{\sigma_\textrm{ISI}}{\mu_\textrm{ISI}}.
	\end{equation}
    In order to compute it, the inter-spike intervals for each individual neurons where obtained. After the CV is computed for each neuron, one can estimate the real CV by averaging over all the nodes.

	\emph{Cross-correlation. (CC)} Given two time series $e(t)$ and $i(t)$ the normalized cross-correlation, also called Pearson correlation coefficient is defined as:

	\begin{equation}
		\text{CC}(\tau)= \tilde \rho_e(t) \tilde \rho_i(t+\tau), 
	\end{equation}

	where $\tilde \rho_{e,i}$ represents a normalized time series substracting the average and dividing by standard deviation, $\tilde \rho(t) = \langle \left(\rho(t) - \langle\rho\rangle  \right) \rangle / \sigma(\rho)$. 	Defined in this way, if CC$(\tau)$ has a peak for $\tau>0$, we conclude that the activity of the inhibitory population resembles that of the excitatory one, but delayed: excitatory population spikes first and it is followed by the inhibitory one.

	\emph{Excitability. } We measure the excitability of the system, i.e., its capability to exhibit long transient behaviours and large amplification of perturbations while being stable, by means of different observables:
	\begin{enumerate}[(i)]
		\item The Henrici index, defined as
		\begin{equation} \label{eqn:Henrici}
			\mathcal{NN}(A)=\sqrt{||A||_F-\sum_{i}^{n}|\lambda_i|^2}, 
		\end{equation}
		where $||A||_F\equiv\sqrt{\sum_{ij}^n |A_{ij}|^2}$ is the Frobenius norm and $\lambda_i$ are the eigenvalues of $A$. This index, that measures the non-normality of a matrix, allows us to estimate the excitability of our system directly from its Jacobian matrix (the exact relation between non-normality and excitability is clarified below).  
		\item Direct stimulation of the system by externally activating a fraction of the excitatory neurons. This method allows one to visualize the degree of excitability of the model in sparse networks where the solutions cannot be calculated analytically. 
		
	\end{enumerate}

	\subsection{Non-normality, Non-reciprocity, reactivity and excitability \label{sec:reactivity}}
	Through this article we employ the concepts of non normality, non reciprocity and excitability in similar contexts. In this section we aim to clarify the relations existing between such concepts.
	
	\begin{itemize}
		\item A matrix $A$ is \emph{normal} if and only if $AA^*=A^*A$, i.e, it commutes with its transpose conjugate $A^*$. Equivalently, $A$ is normal if it has a complete set of orthogonal eigenvectors, that is, if it is unitarily diagonalizable: $A=UDU^*$ \cite{PseudoSpectra}. Therefore, \emph{non-normal} matrices are those that do not commute with its transpose conjugate or equivalently, do not have orthonormal eigenvectors. A simple measure of non-normality is given by the Henrici index defined in Eq.(\ref{eqn:Henrici}), although other possibilities for quantifying non-normality exist in the literature \cite{NonNormalityMeasures}.
		\item A system is \emph{non-reciprocal}, when interactions between its components are asymmetrical, i.e. the way agent $a$ interacts with agent $b$ is different from the way $b$ interacts with $a$. In the context of dynamical systems, a system is non-reciprocal if the Jacobian has different off-diagonal entries.
		\item \emph{Excitability} is defined as the capability of a system to exhibit long transient behaviours (i.e. large trajectories away from its stationary state) and large amplification of perturbations.
		\item \emph{Reactivity} is a simple way to quantify the magnitude of the transient behaviour of a dynamical system before relaxing to its stationary state. In the literature of non-normal matrices it is also known as the numerical abscissa of a matrix, which in the case of the Hilbert space is given by \cite{PseudoSpectra}:
		\begin{equation}
			\mathcal{R}(A)=\lambda_\text{max}(H(A)),
		\end{equation}
		where $A$ is a matrix and $\lambda_\text{máx}(H(A))$ is the maximum eigenvalue of the Hermitian part of $A$ defined as $H(A)=\frac{1}{2}(A+A^*)$. Although it may then seem more natural (at least \emph{a priori}) to consider reactivity of the Jacobian as a measure of its excitability instead of non-normality, we demonstrate below that for two dimensional matrices, reactivity can be increased either by decreasing the stability of the fixed point or by increasing the non normality of the Jacobian. Thus, non normality is precisely what makes a state with fixed stability more excitable.
		
	\end{itemize}
		Let us illustrate all these concepts with the most simple case, a dynamical system with associated Jacobian matrix given by
		\begin{equation}
			J = 
			\begin{pmatrix}
				a  & b \\
				c & d 
			\end{pmatrix}
		\end{equation}
		at one fixed point. We remark that non-reciprocity implies $b\neq c$. The eigenvalues are
		\begin{equation}
			\lambda_\pm=\frac{1}{2}(a+d\pm\sqrt{(a-d)^2+4bc}).
		\end{equation}
		
		Reactivity can also be computed easily, as
		\begin{equation}\label{eqn:Reactivity2D}
			\mathcal{R}(J)=\lambda_\text{max}(H(J))=\frac{1}{2}(a+d+\sqrt{(a-d)^2+(b+c)^2}).
		\end{equation}
		
		The Henrici index reads
		\begin{equation}
			\mathcal{NN}(J)^2=||J||_F-\sum_{i}^{n}|\lambda_i|^2=a^2+b^2+c^2+d^2-2\left(\left(\frac{a+d}{2}\right)^2+\left(\frac{\sqrt{(a-d)^2+4bc}}{2}\right)^2\right),
		\end{equation}
		where the expressions from $\lambda_\pm$ has been used. In order to continue it is important to distinguish if eigenvalues are real or imaginary. For real eigenvalues ($(a-d)^2>-4bc$), the square of the square root can be readily simplified, 
		\begin{equation}
			\begin{split}
			\mathcal{NN}(J)^2=(b-c)^2.
			\end{split}
		\end{equation}
	
		Thus, for a stable fixed point with fixed stability, i.e., $\text{Re}[\lambda_+]=\frac{1}{2}(a+d+\sqrt{(a-d)^2+4bc})=\text{const}<0$, its reactivity is entirely determined by non-normality. Indeed, rewriting Eq.(\ref{eqn:Reactivity2D}) conveniently,
		\begin{equation}
			\mathcal{R}(J)=\frac{1}{2}(a+d+\sqrt{(a-d)^2+4bc+(b-c)^2}),
		\end{equation}
		all of the terms are fixed by the condition of fixed stability except $(b-c)^2$ which is precisely the amount of non-normality.

		Likewise, for complex eigenvalues ($(a-d)^2<-4bc$):
		\begin{equation}
			\begin{split}
				\mathcal{NN}(J)^2=(a-d)^2+(b+c)^2.
			\end{split}
		\end{equation}
	
		Again fixing the stability of the fixed point, $\text{Re}[\lambda_+]=\frac{1}{2}(a+d)=\text{const}<0$, we have
		\begin{equation}
			\mathcal{R}(J)=\frac{1}{2}((a+d)+\mathcal{NN}(J)),
		\end{equation}
		demonstrating again that for fixed stability, being more non-normal implies also being more reactive.

	%\bibliography{Bib-EICP-PRL}